\documentclass[aps,pra,twocolumn,floatfix,amsmath,nofootinbib,superscriptaddress,amssymb,bibnotes,noeprint]{revtex4-1}

\usepackage{graphicx}
\usepackage{bm}
\usepackage{epstopdf}
\usepackage{siunitx}
\usepackage{braket}
\usepackage{pbox}
\usepackage[colorlinks, linkcolor = blue, citecolor = blue, filecolor = black, urlcolor = blue]{hyperref}
\usepackage{xspace} 
\usepackage{units}
\usepackage{amsmath,amssymb}
\usepackage{color}
\usepackage[dvipsnames]{xcolor}

\DeclareSIUnit\wn{\cm\tothe{-3}}
\DeclareSIUnit\gauss{G}
\newcommand{\rs}{\rm 
\scriptscriptstyle}
\newcommand{\OD}{\text{OD}}
\def\nn{\nonumber}
\def\\,MHz{\text{\,MHz}}

\def\beqa{\begin{eqnarray}}
\def\eeqa{\end{eqnarray}}
\def\beq{\begin{equation}}
\def\eeq{\end{equation}}
\def\mum{\ensuremath{\mu}\text{m}\xspace }

\newcommand{\todoc}[1]{[TODO: \textcolor{gray}{\footnotesize	  #1} ]}
\newcommand{\cmntc}[1]{[CmntC: \textcolor{black}{ \footnotesize #1} ]}
\renewcommand{\cmntc}[1]{}
\renewcommand{\todoc}[1]{}
\newcommand{\remove}[1]{}

\newcommand{\changed}[2]{#1}
\newcommand{\integral}[1]{\int \! \mathrm{d} #1\,}    
\newcommand{\parajcite}[1]{\cite{#1}}

\begin{document}

\title{Photon propagation through dissipative Rydberg media at large input rates}
\date{\today}

\author{Przemyslaw Bienias}
\affiliation{Joint Quantum Institute, National Institute of Standards and Technology and the University of Maryland, College Park, Maryland 20742 USA}

\author{James Douglas}
\affiliation{ICFO-Institut de Ciencies Fotoniques, The Barcelona Institute of Science and Technology, 08860 Castelldefels, Barcelona, Spain}

\author{Asaf Paris-Mandoki}
\affiliation{Department of Physics, Chemistry and Pharmacy, Physics@SDU, University of Southern Denmark, 5320 Odense, Denmark}
\affiliation{Instituto de F\'{i}sica, Universidad Nacional Aut\'{o}noma de M\'{e}xico, Mexico City, Mexico}

\author{Paraj Titum}
\affiliation{Joint Quantum Institute, National Institute of Standards and Technology and the University of Maryland, College Park, Maryland 20742 USA}
\affiliation{Joint Center for Quantum Information and Computer Science, National Institute of Standards and Technology and the University of Maryland, College Park, Maryland 20742 USA}

\author{Ivan Mirgorodskiy}
\affiliation{Department of Physics, Chemistry and Pharmacy, Physics@SDU, University of Southern Denmark, 5320 Odense, Denmark}

\author{Christoph Tresp}
\affiliation{Department of Physics, Chemistry and Pharmacy, Physics@SDU, University of Southern Denmark, 5320 Odense, Denmark}

\author{Emil Zeuthen}
\affiliation{Niels Bohr Institute, University of Copenhagen, DK-2100 Copenhagen, Denmark}

\author{Michael J. Gullans}
\affiliation{Joint Quantum Institute, National Institute of Standards and Technology and the University of Maryland, College Park, Maryland 20742 USA}
\affiliation{Joint Center for Quantum Information and Computer Science, National Institute of Standards and Technology and the University of Maryland, College Park, Maryland 20742 USA}
\affiliation{Department of Physics, Princeton University, Princeton, New Jersey 08544 USA}

\author{Marco Manzoni}
\affiliation{ICFO-Institut de Ciencies Fotoniques, The Barcelona Institute of Science and Technology, 08860 Castelldefels, Barcelona, Spain}

\author{Sebastian Hofferberth}
\affiliation{Department of Physics, Chemistry and Pharmacy, Physics@SDU, University of Southern Denmark, 5320 Odense, Denmark}

\author{Darrick Chang}
\affiliation{ICFO-Institut de Ciencies Fotoniques, The Barcelona Institute of Science and Technology, 08860 Castelldefels, Barcelona, Spain}

\author{Alexey V. Gorshkov}
\affiliation{Joint Quantum Institute, National Institute of Standards and Technology and the University of Maryland, College Park, Maryland 20742 USA}
\affiliation{Joint Center for Quantum Information and Computer Science, National Institute of Standards and Technology and the University of Maryland, College Park, Maryland 20742 USA}

\begin{abstract}
We study the dissipative propagation of quantized light in interacting
Rydberg media under the conditions of electromagnetically induced transparency (EIT).
Rydberg blockade physics in optically dense atomic media leads to strong dissipative interactions between single photons.
The regime of high incoming photon flux constitutes a challenging many-body dissipative problem. We experimentally study in detail for the first time the pulse shapes and the second-order correlation function of the outgoing field and compare our data with simulations based on two novel theoretical approaches well-suited to treat this many-photon limit.
At low incoming flux, we report good agreement between both theories and the experiment. 
For higher input flux, the intensity of the outgoing light is lower than that obtained from theoretical predictions. We explain this discrepancy using a simple phenomenological model taking into account pollutants, which are nearly-stationary Rydberg excitations coming from the reabsorption of scattered probe photons. 
At high incoming photon rates, the blockade physics results in unconventional shapes of 
measured correlation functions.
\end{abstract}

\maketitle

\section{Introduction}\label{man:intro}
A number of platforms enable strong interactions between photons at the level of single quanta~\cite{Chang2014}, 
with Rydberg electromagnetically induced transparency (rEIT)~\cite{Lukin2001,Friedler2005} 
being particularly promising
\cite{Hofmann2013,Mohapatra2007,Parigi2012,Dudin2012,Maxwell2013,Peyronel2012}. 
Rapid progress in the control of rEIT at the level of a few photons has led to the demonstration of strong interactions 
\cite{Maxwell2013,Peyronel2012,Firstenberg2013,Busche2017,Jia2018,Tiarks2016,Thompson2016},
a single-photon source 
\cite{Dudin2012},
atom-photon entanglement 
\cite{Li2013},
a single-photon switch~\cite{Baur2014}, a~transistor~\cite{Gorniaczyk2014,Tiarks2014,Gorniaczyk2016},
and three-body interactions~\cite{Gullans2016,Jachymski2016,Gullans2017,Liang2017}. 
Due to the high tunability and strong interactions offered by rEIT, exotic states of light such as different types of bound states~\cite{Firstenberg2013,Bienias2014,Maghrebi2015c,Gullans2017,Moos2017,Liang2017}, as well as
Wigner crystals of individual photons~\cite{Otterbach2013,Moos2015}, have been predicted and experimentally demonstrated  \cite{Firstenberg2013,Liang2017}.

Generally, however, because of the many-body nature of the underlying open quantum system, the problem of strongly interacting photons is challenging to solve. Brute-force analytical or numerical approaches thus far remain limited to three or fewer 
photons~\cite{Firstenberg2013,Liang2017,Moos2017}. In recent years, progress has been made to develop effective theories for strongly interacting Rydberg polaritons in 1D. These theories are expected to be valid in various dispersive~\cite{Otterbach2013,Bienias2014,Bienias2016,Gullans2016}  and dissipative~\cite{Gorshkov2013,Zeuthen2017} regimes. 
Additionally, a promising numerical algorithm has emerged, based upon a matrix-product-state ansatz and the input-output formalism~\cite{Caneva2015,Shi2015,Manzoni2017}. 
Here, we show that the effective and numerical methods presented in Refs.~\cite{Zeuthen2017} and~\cite{Manzoni2017}, respectively, enable quantitative comparisons with rEIT experimental results, and together  provide new insights into the microscopic workings of these experiments.

After the demonstration of dissipative Rydberg 
  blockade at the single-photon level~\cite{Peyronel2012,Maxwell2013,Boddeda2016}, a natural next step is the realization of a regular train of single photons, which could find many applications in quantum information and metrology~\cite{Motes2014,Zoller2005,Brida2010}. 
Here, we address this timely and exciting problem both theoretically and experimentally.

\begin{figure}[h!]
\begin{center}
\includegraphics[width= 1\columnwidth]{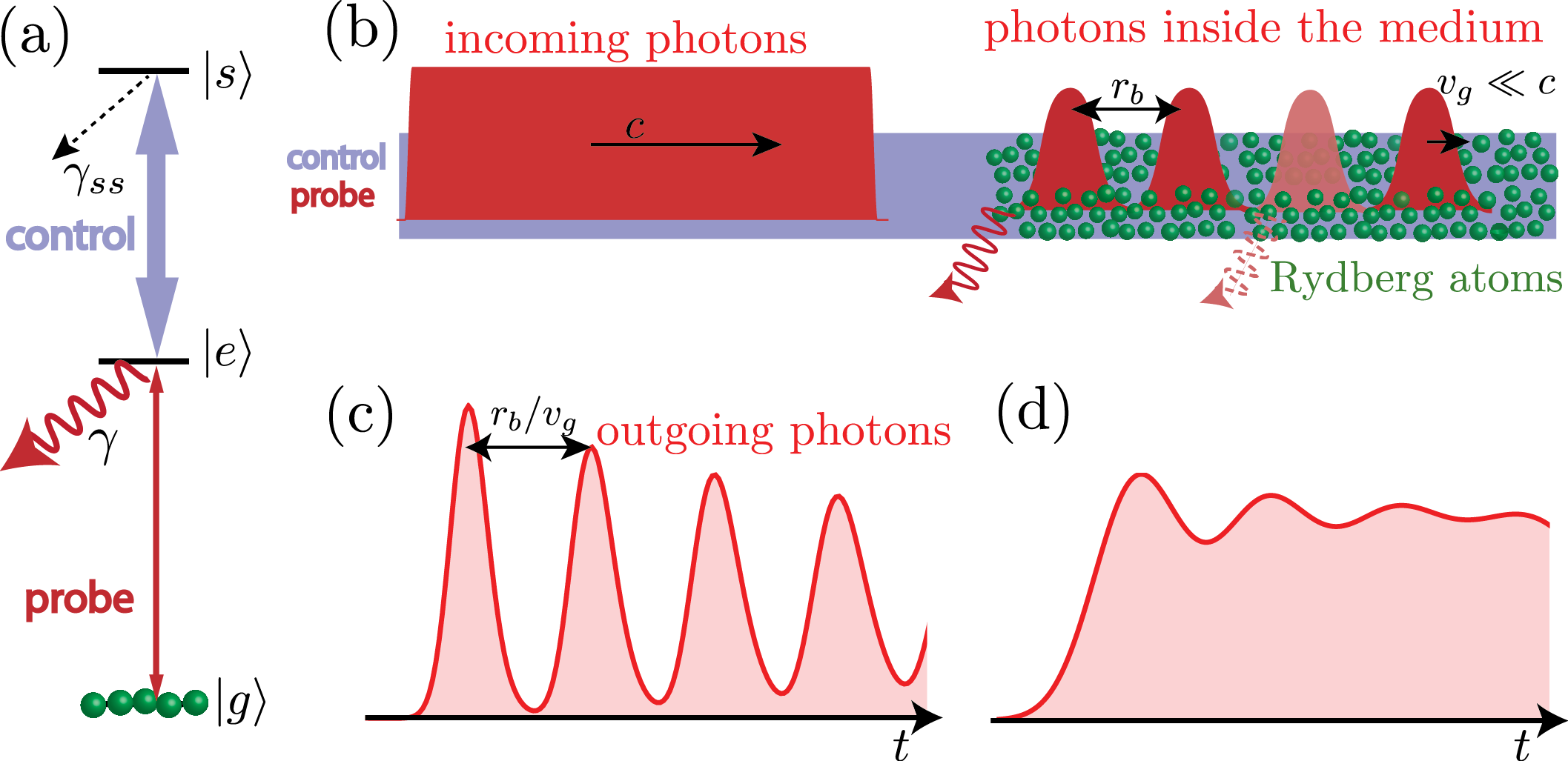}
\caption{\label{system} 
(a) Three-level rEIT scheme, where $\ket{g}$, $\ket{e}$, and $\ket{s}$ are ground, excited, and long-lived Rydberg states, respectively. State $\ket{e}$ decays spontaneously at rate $\gamma$, while $\gamma_{ss}$ describes the decoherence of $\ket{s}$. 
(b) Schematic representation of the theoretical model in position space at two time instances. A photon pulse, incident on the medium with velocity $c$, propagates as Rydberg polaritons with a group velocity $v_g$ inside the medium. However, due to Rydberg blockade, only one polariton per $r_b$ can propagate without loss. All other photons are scattered (represented by the solid wavy red arrow) at the beginning of the medium whenever a polariton is already inside the medium within $r_b$. 
There are additional losses in the medium (represented by the dashed wavy red arrow) due to the finite width of the EIT transparency window.
(c-d) Output pulse shapes as a function of time predicted by the theory (see Sec.~\ref{sec:Theory}), for two different choices of incoming rates $R_{\rs in}$, blockade time $\tau_b$, and EIT filtering time $\tau_{\rs EIT}$:
(c) A time trace for $\tau_{\rs EIT}=\tau_b/5$ and $R_{\rs in}=10/\tau_b $, which gives rise to a train of photons~\cite{Zeuthen2017}. 
(d) A time trace for $\tau_{\rs EIT}=\tau_b/2$ and $R_{\rs in}=3/\tau_b $, which are closer to the parameters accessible in current experiments and in this work. Instead of well separated humps, the intensity  exhibits oscillations with the peaks corresponding to the photon humps in (c). 
}
\end{center}
\end{figure}
%
To be more specific, we consider photons propagating through a Rydberg medium, Fig.~\ref{system}, 
in the regime in which a probe field $\mathcal{E}$ is on resonance with the $\ket{g}-\ket{e}$ transition -- the so-called dissipative regime~\cite{Murray2016,Firstenberg2016}. Van der Waals interactions between Rydberg levels lead to a blockade effect, where effectively only one atom may be excited to the Rydberg level $\ket{s}$ within a blockade radius $r_b$. The remaining atoms within the blockade radius then act as two-level atoms scattering incoming light. In the limit of large optical depth per blockade radius $\OD_b=\OD r_b/L$ (where $\OD$ is the total optical depth for a medium of length $L$), only  one photon per $r_b$ can enjoy EIT and propagate through the medium without loss, while other photons are scattered at the beginning of the medium [depicted by the solid wavy red arrow in Fig.~\ref{system}(b)]. 
For high enough incoming rates $R_{\rs in}\gg 1/\tau_b$ (where $\tau_b=r_b/v_g$ is the blockade time and $v_g$ the EIT group velocity in the medium), a probe pulse shape with a well-defined beginning (sharp enough)  
can give rise to a train of single photons. 
The basic idea behind this train of photons is as follows. The first photon at the \changed{leading edge}{beginning} of the pulse forms a polariton in the beginning of the medium, $r=0$, while a second  photon 
can enter the medium only after the first polariton has propagated at least $r_{b}$ into the medium, $r>r_b$. 
Hence, for higher $R_{\rs in}$, there is a high probability that one or more photons are scattered at the beginning of the medium leading to a projective measurement of the position of the polariton inside the medium, 
making this polariton shorter in time and hence wider in frequency.
Due to the finite width of the EIT transparency window, these high-bandwidth  polaritons~\cite{Zeuthen2017} can decay in the medium  [depicted by the dashed wavy red arrow in Fig.~\ref{system}(b)],
which puts additional constraints on $\OD_b$ and $\OD$ required to observe an outgoing train of single photons~\cite{Zeuthen2017}. 

In this work, for the first time, we experimentally demonstrate the time traces and correlation functions of the transmitted field in the regime of high incoming photon intensity and strong interactions. 
Up to now, Rydberg blockade physics in the dissipative regime resulted in the study of antibunching for photons separated by times smaller than the blockade time $\tau_b$,  $|t|<\tau_b$.
Here, we show experimentally and explain theoretically qualitatively new 
signatures of the blockade in 
the two-photon correlation function $g^{(2)}(\tau)$ as well as in the time traces $R(t)$. In particular, Rydberg blockade leads to a local maximum in $R(t)$ and $g^{(2)}(\tau)$ outside the blockade time $\tau_b$. This hump in output intensity [shown schematically in Fig.~\ref{system}(d)] and  correlations comes from the interplay of blockade physics, the finite width of the EIT transparency window, and the temporal shape of the input pulse.
With this in mind, 
we extend the serialized hard-sphere model introduced in Ref.~\cite{Zeuthen2017} to include the temporal shape of the incoming photons as well as the decoherence of the Rydberg level. We show good agreement with output time traces predicted from exact numerics based on matrix product states (MPS)~\cite{Manzoni2017}. We explore this regime experimentally and find qualitative signs of what the theories predict. Both the theoretical model and MPS numerics differ quantitatively from the experimentally observed time traces and correlations for high incoming photon rates. We believe that these deviations between theory and experiment are due to Rydberg pollutants, i.e.,  additional Rydberg excitations (in $\ket{s}$ and other nearby Rydberg states) which are created by scattered probe photons. 
In order to capture the effect of pollutants, we describe a simple toy model for the dynamics of the pollutants in the system. These pollutants also prevent us, 
as well as other rEIT experiments, from seeing multiple subsequent humps in correlation functions [Fig.~\ref{system}(c-d)]. In particular, pollutants prevent us from accessing higher rates for which humps would be more pronounced and therefore would lead to an output train of single photons. 

The remainder of the paper is organized as follows. In Sec.~\ref{sec:Theory}, we present
two modeling approaches describing dissipative Rydberg EIT at large input rates.
We first present a hard-sphere serialized model, then a model based on  matrix product states, and finally compare their predictions.
In Sec.~\ref{man:experiment}, we present experimental results, compare them with the theory, and discuss measurements suggesting that in order to explain observed data we need to include pollutants.
In Sec.~\ref{pollutants}, we explain in detail the source and impact of the pollutants, as well as describe a numerically tractable toy model capturing the relevant physics. 
This leads to the quantitative agreement between the theory and the experiment.
We summarize our work and give an outlook in Sec.~\ref{sec:outlook}.


\section{Theory of dissipative Rydberg EIT} \label{sec:Theory}
The propagation of resonant light through a medium depends on the level structure of the atoms constituting the medium. In particular, a resonant two-level medium with levels $|g\rangle$ and $|e\rangle$ yields exponential attenuation of the transmission intensity by a factor $T=\exp(-\OD)$. Adding a third level $|s\rangle$ and an appropriate control field makes the medium transparent, as interference suppresses population in $|e\rangle$, and a dark-state polariton with slow group velocity is generated. This effect is known as electromagnetically induced transparency (EIT)~\cite{Fleischhauer2005}. 

Let us now consider the propagation of photons through a dense medium of interacting three-level atoms 
under EIT conditions. Fig.~\ref{system}(a) shows the level structure of the atoms with levels $|g\rangle$, $|e\rangle$, and $|s\rangle$. The control field has a Rabi frequency $\Omega$ [it takes time $\pi/\Omega$ to do a $\pi$ pulse], and $\gamma$ is the fullwidth of the level $|e\rangle$.  The output intensity can be calculated using the following theory for dissipative dynamics developed in Refs.~\cite{Gorshkov2011,Gorshkov2013,Zeuthen2017}. A single photon incident under EIT conditions is converted into a Rydberg polariton (approximately a Rydberg spin wave) 
moving at a reduced group velocity $v_g$. In the presence of strong Rydberg-Rydberg van der Waals interactions of the form $C_6/r^6$, this Rydberg polariton destroys EIT for any subsequent photon incident within a blockade radius 
$r_b=\left[C_6\left(\frac{1}{2\gamma_{\rm EIT}}+\frac{1}{\gamma}\right)\right]^{1/6}$, where $\gamma_{\rm EIT}=\frac{\Omega^2}{2\gamma}$ is the single-atom EIT linewidth~\cite{Bienias2014}. 
In the limit of large blockaded optical depth $\OD_b=\OD\frac{r_b}{L}$, this leads to strong dissipation and absorption of all photons incident within a blockade time, $\tau_b=r_b/v_g=\OD_b/(2\gamma_{\rs EIT})$, after the formation of a polariton. This is shown schematically  in Fig.~\ref{system}(b). 
Ideally, this would lead to the conversion of a continuous-wave input into a train of single photons separated in free space by the decompressed blockade radius, $r_b c/v_g$. 
However, the propagating polariton may decay because of the finite width of the EIT window, which washes out any spectral features sharper than $1/\tau_{\rm EIT}=\gamma_{\rm EIT}/\sqrt{\OD}$. 
Based on this intuition, the approximate output intensity 
[within the so-called hard-sphere model] 
may be obtained through a serialized approach in which we first determine the output due to dissipative Rydberg-Rydberg interactions for perfect single-polariton EIT conditions, and then frequency-filter the output with a filter of width $1/\tau_{\rm EIT}$~\cite{Zeuthen2017}. 
In contrast, the exact simulation using MPS does not rely on such an ansatz in treating the single-polariton EIT physics. We will now describe the hard-sphere and the MPS models in more detail. 

\subsection{Hard-sphere serialized model \label{subsec:hard}}
In Ref.~\parajcite{Zeuthen2017}, Zeuthen et al.\ develop a hard-sphere model to  calculate outgoing photon rates and pulse shapes for incoming photon pulses that are longer than the medium. The basic assumptions in this model are as follows. Rydberg interactions are approximated by a hard-sphere potential of size $r_b$. The medium is considered to be homogeneous with sharp boundaries, and polaritons only form in the beginning of the medium. Under these assumptions, it is possible 
to compute the output  photon rate and the output time trace for a Poisson-distributed input at constant average input photon rate. 
Throughout the manuscript, 
by the \textit{time trace} we mean the \textit{ensemble averaged time trace}, i.e., the average over many experimental realizations of time traces.
At perfect EIT, because of the hard-sphere dissipative interactions, the output rate for increasing incoming rate is saturated by one photon per blockade time. 
The finite EIT window can be accounted for by considering the effect of the scattered photons: Once the first polariton is formed at the beginning of the medium, the next photon arriving within a blockade time $\tau_b$ of formation is scattered. 
This projects and localizes the first polariton wavefunction, with the time-width of the polariton being determined by the timing of the first scattering event. 
This means that higher input rates of photons will make  the polariton wavefunction more localized in time. 
If the narrow polaritons do not fit in the EIT window [given by $1/\tau_{\rm EIT}$], they may decay. 
This decay is governed by single-polariton physics, and we account for the EIT losses by using a Gaussian filter~\cite{Zeuthen2017}. 
This model was shown to be accurate in the limit of large $\OD_b$, where the predicted transmission rate was compared with exact numerical simulation of two-photon dynamics~\cite{Zeuthen2017}. We review the 
details of this approach in Appendix~\ref{sec:Emil_theory}.

 The model discussed above assumes a constant input rate. Here, we extend this model to account for arbitrary input pulse shapes. We consider the input photons to be well-described by a coherent state, and the temporal shape is given by a real envelope $h(t)$ satisfying $\integral{t} h^2(t)=1$. For the sake of brevity, we relegate technical details to the Appendix~\ref{sec:Emil_theory}. 
 The Tukey pulse shape $h(t)$, which is used in the experiment, is shown schematically in Fig.~\ref{pulse}(b), where there is a ramp over the time $t_{\rm rise}$ followed by a constant input rate. 
We first calculate the intensity $G^{(1)}(t)=\left\langle\mathcal{E}^\dagger (t)\mathcal{E}(t)\right\rangle$ taking into account only blockade  without EIT filtering.
Then, we calculate the off-diagonal correlation function $G^{(1)}(t,t')=\left\langle\mathcal{E}^\dagger (t)\mathcal{E}(t')\right\rangle$ and express it in terms of intensities $G^{(1)}(t)$. 
Finally, we convolve $G^{(1)}(t,t')$ with a Gaussian filter function, which enables us to estimate the effect of a finite EIT window and leads to the intensity profiles shown in Fig.~\ref{MPSuniform}.

In the regime of $t_{\rs rise} \ll \tau_b$, the output intensity predicted by the hard-sphere model is a train of single photons only in the limit of large input rates $R_{\rm in}\gg1/\tau_b$ and large $\OD_b$. In terms of timescales, the condition on large $\OD_b$ corresponds to a large blockade time, $\tau_b\gg \tau_{\rm EIT}$. Physically, this condition means that the photons in the train, which are necessarily each shorter than $\tau_b$, fit into the EIT transparency window, which has width $1/\tau_{\rm EIT}$. We define the ratio of these two time scales, $\nu=\tau_b/\tau_{\rm EIT}=\OD_b/2\sqrt{\OD}$ as a parameter quantifying whether it is possible to observe the photon train. In this scenario, $G^{(1)}(t)$ would exhibit pronounced oscillations as a function of time, as shown in Fig.~\ref{system}(c), where we plot the predicted time trace for $\tau_{\rm EIT}=\tau_b/5$, i.e.\ $\nu = 5$. As long as $\nu>1$ and $R_{\rs in} $ is appropriately chosen, the hard-sphere theory predicts oscillations in $G^{(1)}(t)$ with the separation of the peaks approximately given by $\tau_b$. However, in the experimentally relevant regime, we have $\nu\approx 1$. In this case, if we attempt to raise $R_{\rm in}$ above $1/\tau_b$ to obtain the train, any oscillations in $G^{(1)}(t)$ 
are washed out due to strong filtering. 

\begin{figure}
\begin{center}
\includegraphics[width= .9\columnwidth]{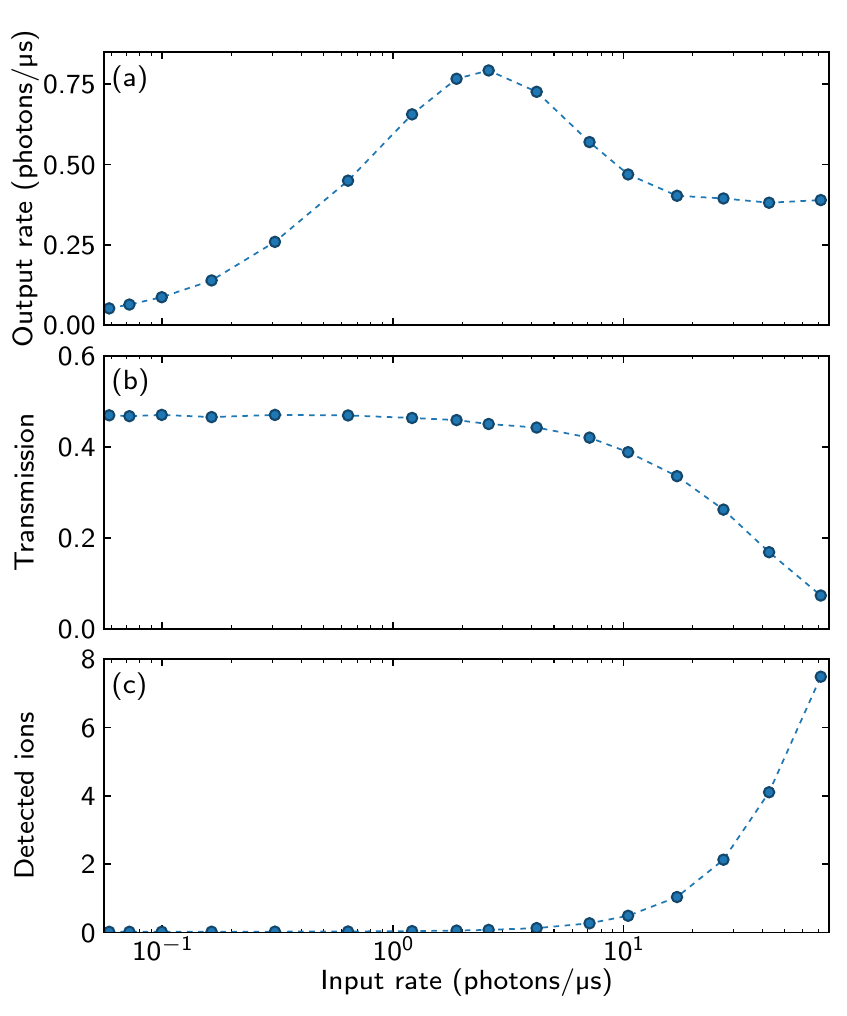}
\caption{\label{rates} 
As a function of the input photon rate, the figure shows (a) output photon rate in the steady-state region of the output pulse [see Fig.~\ref{pulse}(c,d)], (b) transmission of the weak Gaussian test pulse through the medium, and (c) the number of detected ions. Lines connecting points are only a guide to the eye.}
\end{center}
\end{figure}

An interesting feature observed in the predicted time traces of the output intensity (Fig.~\ref{MPSuniform}) is the appearance of a hump at the start of the output time trace for larger input photon rates,
in spite of the strong EIT filtering discussed above.
This hump results from the interplay of two effects present for the parameters and pulse shapes relevant to the experiment:
First, 
the incoming intensity $|h(t)|^{2}$ increases with time which naively would lead to the monotonous increase of the outgoing intensity.
Second, the impact 
of EIT filtering is time-dependent because  it depends on the input photon rate proportional to $|h(t)|^2$;  
and therefore, for greater $h(t)$, each polariton is more localized due to the position-projecting scattering of photons at the beginning of the medium. 
In summary, the interplay of rising incoming intensity and stronger filtering at later times may (and for our parameters does) lead to a maximum in the outgoing intensity  
around the time when the amplitude of the output pulse settles to an approximately constant steady-state value
(i.e., around approximately $t_{\rs rise}+\tau_d$, where $\tau_d$ is the time delay of the transmitted pulse compared to the reference pulse).
For a slower rise of $h(t)$ (i.e., larger $t_{\rs rise}$), the hump gets smaller. Note that the hump in the time trace indicates the existence, for a continuous-wave experiment, of an optimal input photon rate where the outgoing photon rate is maximum. 
One indeed sees a local maximum when plotting the outgoing steady state as a function of the input rate where the interplay between dissipative interactions and EIT gives rise to a hump~\cite{Zeuthen2017}. This is also  consistent with experimental observations [see Fig.~\ref{rates}], however, the involved physics is more complex as we will  discuss in Sec.~\ref{man:experiment}.
\subsection{Detection of Rydberg pollutants \label{ExpPollutants}}

As a corollary, one might consider what happens at the end of the pulse. In this region, the incoming pulse rate $R_{\rm in}(t)$ decays to zero in a time $\sim t_{\rm fall}$. Using the same logic of weaker filtering for smaller intensities, one expects the presence of a hump at the end of the output pulse. However, as we discuss in Sec.~\ref{man:experiment}, the experimental measurements indicate the absence of any such hump at the end of the pulse. This leads us to conjecture the role of pollutants, which explains both the amplification of the hump in the beginning and the lack of a hump at the end of the output pulse. We discuss a simple model for the pollutants and its consequences in Sec.~\ref{pollutants}.

\subsection{MPS method\label{MPSmethod}}
In addition to the hard-sphere model described in the previous Section, we can also numerically obtain the output time traces using a novel time-evolution technique based on MPS introduced in Ref.~\cite{Manzoni2017}. This method, presented in greater detail in Appendix~\ref{sec:MPSappendix}, 
relies on mapping the Maxwell-Bloch equations describing the original atomic ensemble to the propagation of a quantum field through a one-dimensional waveguide coupled to atoms. One key to this mapping is the use of a much smaller number of atoms [$N\lesssim 100$] in the waveguide system (relative to the true number of atoms), while tuning the system parameters to ensure that macroscopic properties such as the optical depth and optical depth per blockade radius remain the same. 
Furthermore, all of the field properties are expressed in terms of the input field and correlation functions of the atoms alone via an input-output relation, while the dynamics of the atoms interacting with the field are encoded in a quantum spin model. As a final step, the dynamics are then solved using the MPS ansatz.
The ansatz relies on the fact that, in many systems, the complete Hilbert space, which grows exponentially with atom number, is not necessary for a faithful representation of the physical states that occur, and, instead, a substantially restricted set of states, those formed from matrix products, is sufficient. This method has been extremely successful in studying condensed-matter many-body problems that would be intractable using direct diagonalization, and in Ref.~\cite{Manzoni2017} was applied to light propagation in atomic ensembles. Here we extend the method in Ref.~\cite{Manzoni2017} to propagate the density matrix of the rEIT system in time, allowing for efficient numerical simulation of the highly dissipative system we study here.

The main benefit of the MPS method is that it allows a quantitative description beyond the hard-sphere model. Specifically, the nature of EIT in the Rydberg system is captured from first principles by using three-level atoms [as shown in Fig.~\ref{system}(a)] directly in the simulation, rather than applying an approximate filter function to the photon wavepacket.
Furthermore the full spatial form of the Rydberg interaction can be approximated to arbitrary precision by a sum of exponential interactions that are efficiently represented within the MPS method. Other details such as inhomogeneity in the atomic cloud, arbitrary time dependence of the input beam, and losses due to spontaneous emission and pure dephasing can also be implemented directly (see Appendix~\ref{sec:MPSappendix}). This allows us to check the results of the more intuitive hard-sphere model and to make qualitative comparisons with experimental results. Furthermore, we expect that this method will also be useful in other regimes of rEIT where effective models are not available.

\subsection{Comparison between the MPS method and the hard-sphere model \label{MPSvsHardsphere}}
{Fig.~\ref{MPSuniform} shows time traces from the MPS and the hard-sphere models for a uniform atomic cloud with all other parameters as in the experiment.
We fix $\OD$ and take $L=47.2\,\mum$. 
%
%
\begin{figure}
\begin{center}
\includegraphics{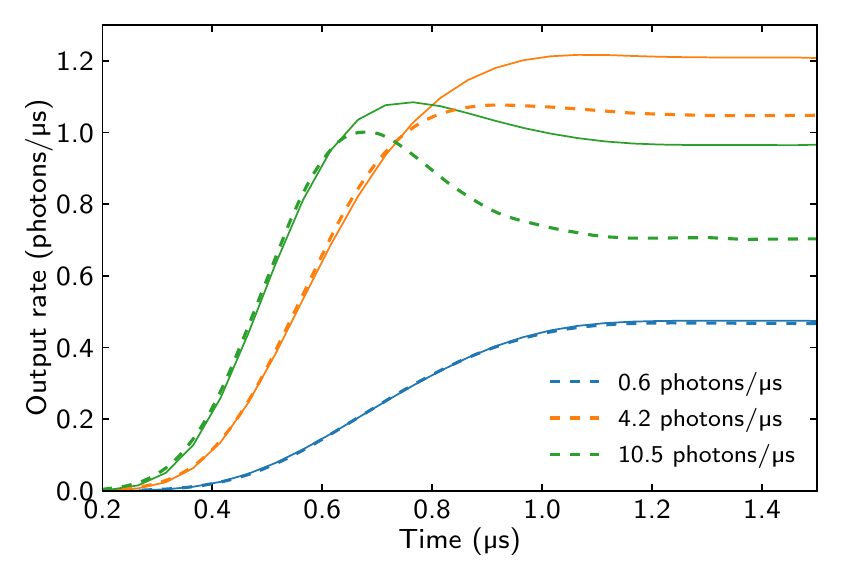}
\caption{\label{MPSuniform} 
Time traces of the output pulse for a uniform cloud with an input Tukey pulse of the same shape as the one used in the experiment. 
Comparison between MPS (dashed lines) and effective hard-sphere model (solid lines) without free parameters. The atomic cloud is taken as uniform in both models with $L = 47.2\,\mu m$, chosen to be consistent with the length of the experimental cloud [described in Sec.~\ref{man:experiment}]. 
Other parameters are $\OD=33$, $\gamma/2\pi = 6.065$\,MHz, $\Omega/2\pi = 10$\,MHz and $C_6/2\pi = 1.87820\times 10^{14}$Hz\,$\mu$m$^6$ for the $n=111$ Rydberg state.
Decoherence in the Rydberg level with full width 
$\gamma_{ss}/2\pi = 40$\,kHz, which we extract from the transmission at low incoming rates. MPS simulations used $N = 60$ and bond dimensions $D = 80, 120$, and $160$ for rates 0.6\,ph/$\mu$s = $0.078/\tau_b$, 4.2\,ph/$\mu$s = $0.55/\tau_b$, and 10.5\,ph/$\mu$s = $1.4/\tau_b$, respectively (here ph stands for photons).
Note good agreement at initial times. Also note that the agreement is better for lower incoming rates than for higher ones. 
We do not see multiple humps because 
the blockade time $\tau_b =0.13\,\mu $s and filtering time $\tau_\textrm{EIT} = \sqrt{\OD}/\gamma_\textrm{EIT} =0.11\,\mu $s are comparable, and because the rise time of the pulse $t_{\rm rise}=0.8\,\mu $s is much greater than $\tau_b$.
}
\end{center}
\end{figure}
We see good agreement between MPS and the hard-sphere model for small rates and/or initial times $t<t_{\rs rise}$. 
For higher rates, both methods agree qualitatively, with MPS giving a more pronounced hump.
Note that, without the use of any fitting parameters, the absolute suppression of the incoming photon rate $R_{\rs out}/R_{\rs in}$ in steady state is predicted by both theories to be at the 10\% level. 
While in this sense the two theories agree well at the order-of-magnitude level, their predictions show appreciable relative deviations as seen in Fig.~\ref{MPSuniform}.
This confirms that we can use the intuitive picture based on the hard-sphere model to explain qualitatively, but not quantitatively, MPS numerics and experimental data.

Note that, due to $t_{\rs rise}\gg \tau_b$, the hump in output time traces 
is mostly due to the non-monotonic relationship between steady-state input and output intensities, postulated
in Ref.~\cite{Zeuthen2017} and discussed in section~\ref{comparisonTheoExp}. 
The visibility of a train of photons depends on $\nu$, which in our case  is $\approx 1$, making only the first hump in the train potentially visible. Furthermore, since $t_{\rs rise}\gg \tau_b$, there is a large uncertainty in when the train actually begins, which further washes out the hump associated with the first photon in the train. As a result, the first photon in the train has only a minor contribution to the experimentally observed hump.

\section{Experiment\label{man:experiment}}

\begin{figure}
\begin{center}
\includegraphics[width= 1\columnwidth]{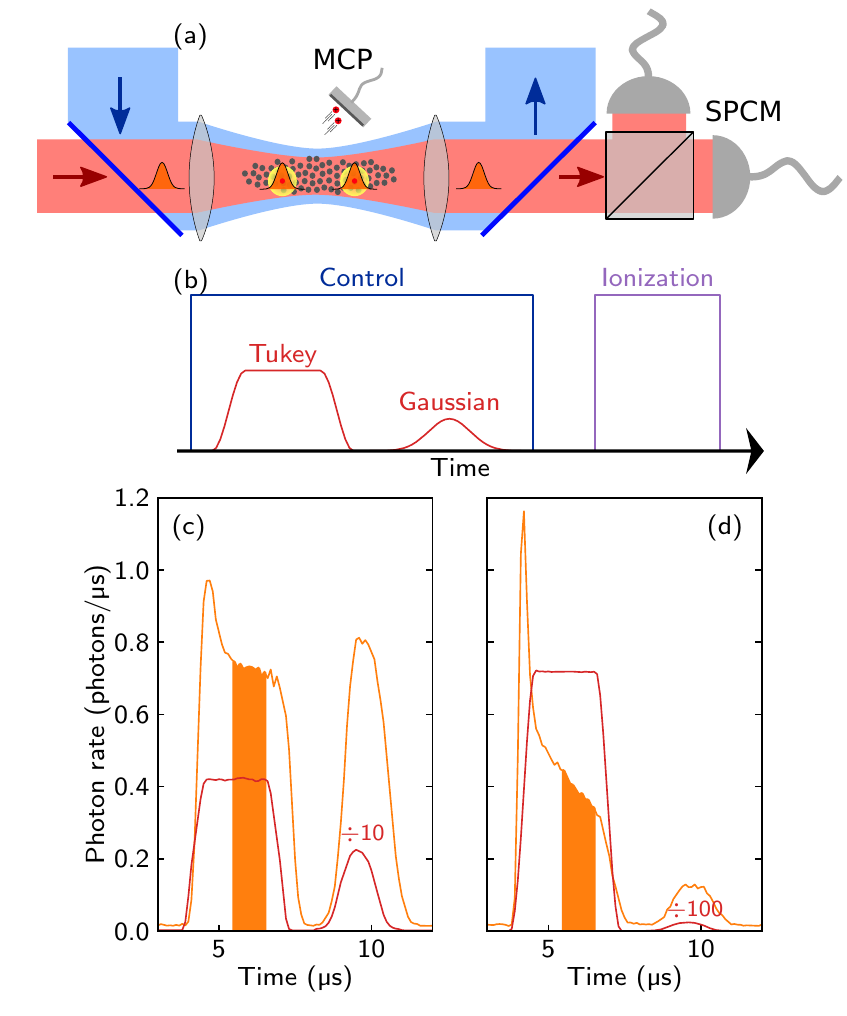}
\caption{\label{pulse}(a) Schematic of the experimental setup showing the probe and control beams focused into the atomic cloud, as well as the detectors for probe photons (SPCMs) and ions (MCP). (b) Illustration of the pulse sequence for a single experimental run. (c) and (d) Output pulse shapes (orange lines) observed in the experiment for input photon rates $R_{\rs in}=4.2$\,ph/$\mu$s = $0.55/\tau_b$ and $R_{\rs in}=71.8$\,ph/$\mu$s = $9.3/\tau_b$, respectively. Red lines depict the input pulses, whose values are divided by factor of 10 in (c) and 100 in (d) for easier viewing. Also shown are the input and output pulses of the weak Gaussian test pulse following the main probe pulse. The main distortion observed in the outgoing probe pulses is the appearance of the initial hump, which becomes more pronounced for higher input photon rates. The orange-shaded regions indicate the timing window we analyze to obtain the steady-state outgoing photon rate.}
\end{center}
\end{figure}
Next, we review the technical details of our experiment.
We start a measurement by preparing $8 \times 10^{4}$ atoms of $^{87}$Rb trapped in an optical dipole trap,  producing a cigar-shaped atomic cloud at \SI{4}{\micro\kelvin} with 
 the density described by $n(z,R)\sim\exp[-R^2/2\sigma_{\rs R}^2-z^2/2\sigma^2]$ where $\sigma_{\rs R}=\SI{6.5}{\micro\meter}$ and $\sigma=\SI{23.6}{\micro\meter}$ characterize radial and longitudinal direction. 
All the atoms are optically pumped into the initial ground state $\Ket{g}=\Ket{5S_{1/2},F = 2, m_F = 2}$.  
We focus a weak \SI{780}{\nano\meter} probe laser beam (Gaussian beam waist $w_{\rs 0,{probe}} = \SI{6.7}{\micro\meter}$) into the cloud [Fig.~\ref{pulse}(a)], coupling the ground state $\Ket{g}$ and the intermediate state $\Ket{e}=\Ket{5P_{3/2}, F = 3, m_F = 3 }$. 
To establish EIT in the system, we add a strong \SI{480}{\nano\meter} control laser beam (Gaussian beam waist $w_{\rs 0,{control}} = \SI{14}{\micro\meter}$) coupling the intermediate state $\Ket{e}$ and the Rydberg state $\Ket{s}=\Ket{111S_{1/2},m_J=1/2}$. The control Rabi frequency is measured to be 
 $\Omega/2\pi = 10\\,MHz$. 
From this, the Rydberg blockade radius is calculated to be $r_{\rs{b}}=\SI{18.7}{\micro\meter}$~\cite{Bienias2014}. For these parameters, we observe a time delay $\tau_d\approx 0.31\,\mu s$ of the weak probe pulses, from which we estimate the optical depth of our medium to be $\OD=33$.

The pulse sequence of a single experimental run is depicted in Fig.~\ref{pulse}(b). To investigate the probe propagation at high photon rates, we send a Tukey-shaped probe pulse (\SI{2}{\micro\second} uptime and \SI{0.8}{\micro\second} rise and fall times) with a varying amplitude into the medium, while the control light is on to maintain EIT conditions. The transmitted probe light is collected on a combination of four single-photon counting modules (SPCMs). Our key experimental observations are the deformation of the probe pulse shapes transmitted through the cloud and the strong dependence of this deformation on the input probe photon rate. Two examples for intermediate and high photon rates are shown in Figs.~\ref{pulse}(c) and (d), respectively. In both cases, we observe the appearance of an initial hump in the transmitted pulses, the width of which is on the order of $\tau_b$. At a very low input photon rate of 0.6\,ph/$\mu$s, this hump is completely absent [Fig.~\ref{fig:comp_exp_MPS}(a)]. At this rate, we only observe weak absorption caused by the Rydberg blockade and by the decoherence of the Rydberg level. We also observe the time delay of the transmitted pulse compared to the reference pulse. Besides the initial hump, we are interested in the steady-state transmission of the outgoing probe light. For this, we consider the 
orange-shaded regions indicated in Figs.~\ref{pulse}(c) and (d), where the transmission becomes approximately constant, as it does over a wide range of input photon rates we measure  (the non-constant nature of this region in Fig.~\ref{pulse}(d) is discussed in the next section).
Fig.~\ref{rates}(a) shows the extracted steady-state transmission of the Tukey pulse as a function of the incoming rate. We find that after reaching a maximum for an input rate of $R_{\rs in}\approx 3$\,ph/$\mu$s, the output photon rate saturates to a constant value [depicted by the orange-shaded time window in Figs~\ref{pulse}(c) and (d)]. Within this time window, we calculate from the experimental data the second-order correlation function $g_2(\tau)$ for the outgoing photons, as shown in Fig.~\ref{corr}. At low input photon rates, we find the previously observed anti-bunching at $\tau=0$ 
caused by the Rydberg-blockade-induced nonlinearity of the medium~\cite{Peyronel2012}. For higher input photon rates, the $g^{(2)}(\tau)$ correlation functions exhibit two striking features. First, the width of the anti-bunching dip shrinks, while at the same time we observe maximal bunching [$g^{(2)}(\tau)>g^{(2)}(0)$]
of photons at separations $\tau$ approximately equal to the blockade time.
Before we compare in Sec.~\ref{comparisonTheoExp} our experimental data to the results of MPS numerics introduced in Sec.~\ref{sec:Theory}, we briefly discuss the experimental observation of Rydberg {pollutants} in the optical medium and how they affect our experiment.

\begin{figure}
\begin{center}
\includegraphics[width= .9\columnwidth]{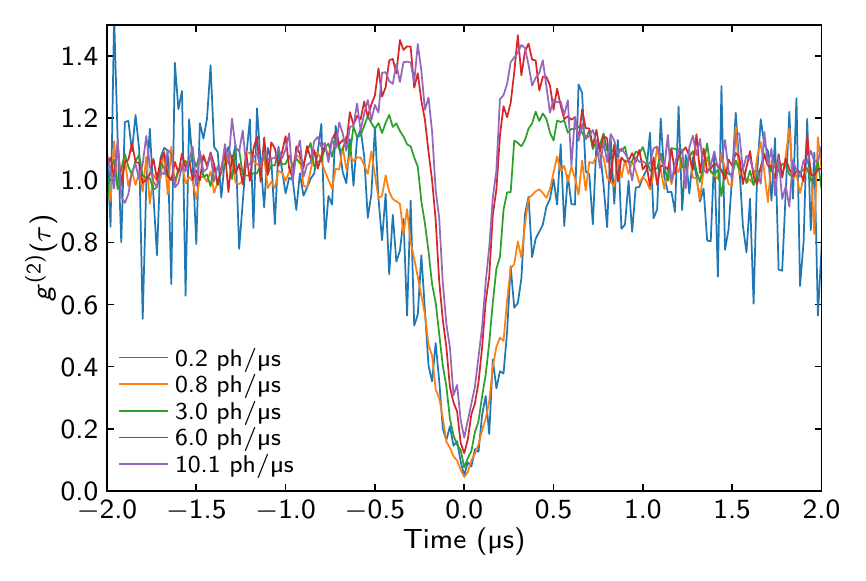}
\caption{\label{corr} 
Second-order correlation function measured for different photon rates between $0.2$\,ph/$\mu$s = $0.026/\tau_b$ and $10.1$\,ph/$\mu$s = $1.3/\tau_b$. 
}
\end{center}
\end{figure}

For the highest photon rates probed in our experiments, we observe that the outgoing probe photon rate, instead of reaching a steady-state value after the initial hump, continues to decrease on a timescale unrelated to the width of the hump [Fig.~\ref{pulse}(d)]. We trace this effect back to the creation of stationary Rydberg excitations that are not accounted for in the theoretical models introduced in Sec.~\ref{sec:Theory}. We quantify the number of these pollutant atoms and their effect on the probe photon transmission in two ways [Fig.~\ref{pulse}(b)]. After each Tukey pulse, we probe the medium with a second Gaussian-shaped probe pulse ($\sigma_{\rs{\tau}}=\SI{0.5}{\micro\second}$ with constant peak amplitude of 2.4\,ph/$\mu$s) to measure how the unwanted Rydberg excitations, created during the Tukey pulse and remaining in the cloud after the initial pulse has passed, reduce the transmission of this weak test pulse. Secondly, after the end of both pulses, we switch off the control light field and immediately ionize any remaining Rydberg atoms. The produced ions are collected on a microchannel plate detector (MCP). The intensity of the Gaussian test pulse is chosen so low and its length so short [compared with the lifetime of the Rydberg states on the order of ms] that the number of detected ions is unchanged by this test pulse.

The two observables characterizing the Rydberg pollutants which we extract from these additional measurements, namely the weak test pulse transmission and the number of detected ions after field ionization,  are shown in Figs.~\ref{rates}(b) and (c), respectively,  as a function of the incoming probe photon rate. Specifically we find that, together with the growing number of detected ions, the test probe pulse transmission is reduced, meaning that the pollutant atoms affect the propagation of probe photons through the polluted medium.

It is important to note that, between the Tukey pulse and the field ionization, the control light is left on for multiple microseconds, which should depump stationary Rydberg excitations created during the probe pulse from the initial $\Ket{s}$ state. The fact that we still find a significant number of ions suggests that these Rydberg atoms have undergone a state change. Our field-ionization voltage is sufficiently high to ionize Rydberg states with $n>50$, ensuring that we ionize atoms over a wide  range of states near the original $\Ket{s}$ state. The claim that many of these atoms have transitioned to a state that interacts with $\ket{s}$ only weakly (or have moved outside of the control and probe beams) is supported by the relatively weak suppression of test pulse transmission they cause. A stationary atom in state $\Ket{s}$ would block a significant part of the atomic cloud ($\OD_b>5 $), resulting in strong attenuation of probe photons. Finally, we notice that the the ion number grows with $R_{\rs in}$ faster than linearly, which suggests that it is, at least partially, a two (or more)-body effect. 

We discuss the possible origins of these pollutants in Sec.~\ref{pollutants}, where we also introduce an effective model to simulate their influence on pulse propagation. 
Fig.~\ref{rates} suggests that this pollution effect becomes significant for large input photon rates $R_{\rs in} >$1 \,ph/$\mu$s. 
To further quantify when this pollution becomes important, in the following section, we compare our experimental observations to the MPS theory developed in Sec.~\ref{sec:Theory}.


\subsection{Comparison of theory and experiment \label{comparisonTheoExp}}
\begin{figure}[t]
\begin{center}
\includegraphics[width= 1\columnwidth]{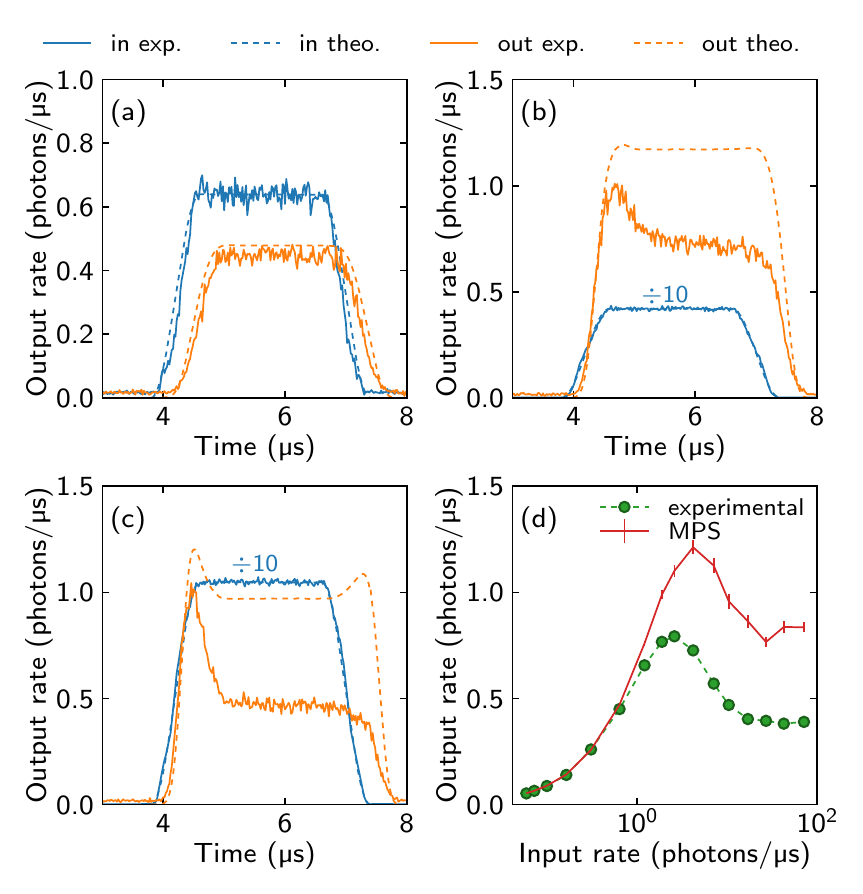}
\caption{Comparison of the experimental photon output with the output simulated using MPS. (a)-(c) Time traces for various input photon rates: (a) 0.6\,ph/$\mu$s = $0.078/\tau_b$, (b) 4.2\,ph/$\mu$s = $0.55/\tau_b$, and (c) 10.5\,ph/$\mu$s = $1.4/\tau_b$. 
From (a), we see that the reference pulse is well described using a Tukey function. 
(d) Steady-state output as a function of the input rate. 
Experimental data and theoretical curves are shown with solid and dashed lines, respectively. Blue curves indicate input pulses while the orange ones depict the output. The input pulses indicated with $\div 10$ have been divided by a factor of 10 for easier viewing. The Rydberg interaction is modeled by a sum of five exponentials as described in Appendix~\ref{ap:expon_approx}. In (a)-(c), MPS density matrix simulations use time step $0.01/\gamma$, $N = 60$ effective atoms, and bond dimensions $D$ equal to (a)-(b) 100, (c) 180. The steady-state results in (d) are from quantum jump MPS simulations with time step $0.01/\gamma$, number of effective atoms $N=70$, and bond dimensions dependent on the input rate as shown in Fig.~\ref{fig:conv_ss} in Appendix~\ref{app:conv_bd_and_ns}.
}\label{fig:comp_exp_MPS}
\end{center}
\end{figure}
\begin{figure}[t]
\begin{center}
\includegraphics[width= .9\columnwidth]{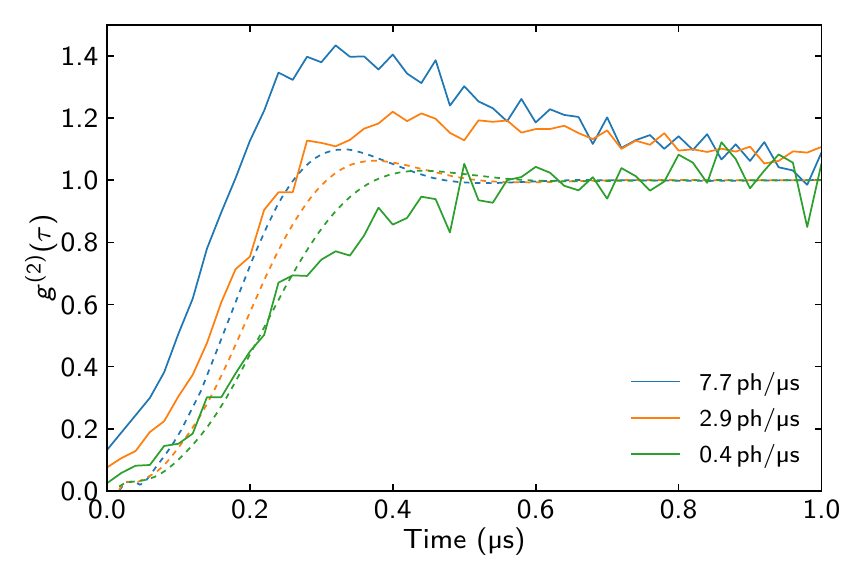}
\caption{\label{g2_comp_mps} 
Experimental second-order correlation function (solid)  compared with its MPS-simulated counterpart (dashed) for different input photon rates. 
Experimental parameters used in MPS simulations are as in all other figures except $\gamma_{ss} /2\pi=120$\,kHz, because the $g^{(2)}$ data was taken in slightly different experimental 
conditions.
The Rydberg interaction is modeled by three exponentials,  as described in Appendix~\ref{ap:expon_approx}. MPS density matrix simulations used time step $0.01/\gamma$, $N = 60$, and bond dimension $D = 140$ for input rates 0.4\,ph/$\mu$s = $0.052/\tau_b$ and 2.9\,ph/$\mu$s = $0.38/\tau_b$, and $D = 180$ for the input rate 7.7\,ph/$\mu$s = $1/\tau_b$.}
\end{center}
\end{figure}

To compare our experimental results quantitatively with theory, we use MPS simulations. The flexibility of the MPS model allows us to treat quantitatively crucial aspects of the experiment, such as the spatial dependence of the Rydberg interaction and the non-uniform cloud density  $n(z)$ along the probe beam direction $z$  
(see Appendix~\ref{sec:MPSappendix} for numerical details). Executing this model with the experimental parameters, 
we show in Fig.~\ref{fig:comp_exp_MPS} the comparisons with the experimental results for time traces at various input rates, as well as the steady-state output rate as a function of input rate.

In the time trace shown in Fig.~\ref{fig:comp_exp_MPS}(a) and for low input rates in the steady state [Fig.~\ref{fig:comp_exp_MPS}(d)], we see excellent agreement 
between the experiment and the MPS model. However, at higher input rates [Fig.~\ref{fig:comp_exp_MPS}(b)-(c) and part of Fig.~\ref{fig:comp_exp_MPS}(d)], we see the presence of a much larger initial hump and lower steady state in the experimental output relative to our numerics. Furthermore, in Fig.~\ref{fig:comp_exp_MPS}(c), a second hump at the end of the output pulse is visible in the MPS simulation, but is absent in the experiment. This suggests that, for these higher rates, the pollution described above plays a 
role in determining both the size of the initial hump in the output pulse and the strength of the steady-state signal.

The pollution  also plays a role in explaining the relation between
$g^{(2)}$ measured in the experiment and the corresponding MPS simulations. In Fig.~\ref{g2_comp_mps}, we show this comparison between the theory prediction and the experimental observations for three different input photon rates. We see that the theory reproduces the qualitative feature of hump size increasing with input rate, however the humps are much larger in the experiment, suggesting once again that pollution is non-negligible at high input rates.  

\section{Pollutants \label{pollutants}}
\begin{figure}[t]
\begin{center}
\includegraphics[width= .8\columnwidth]{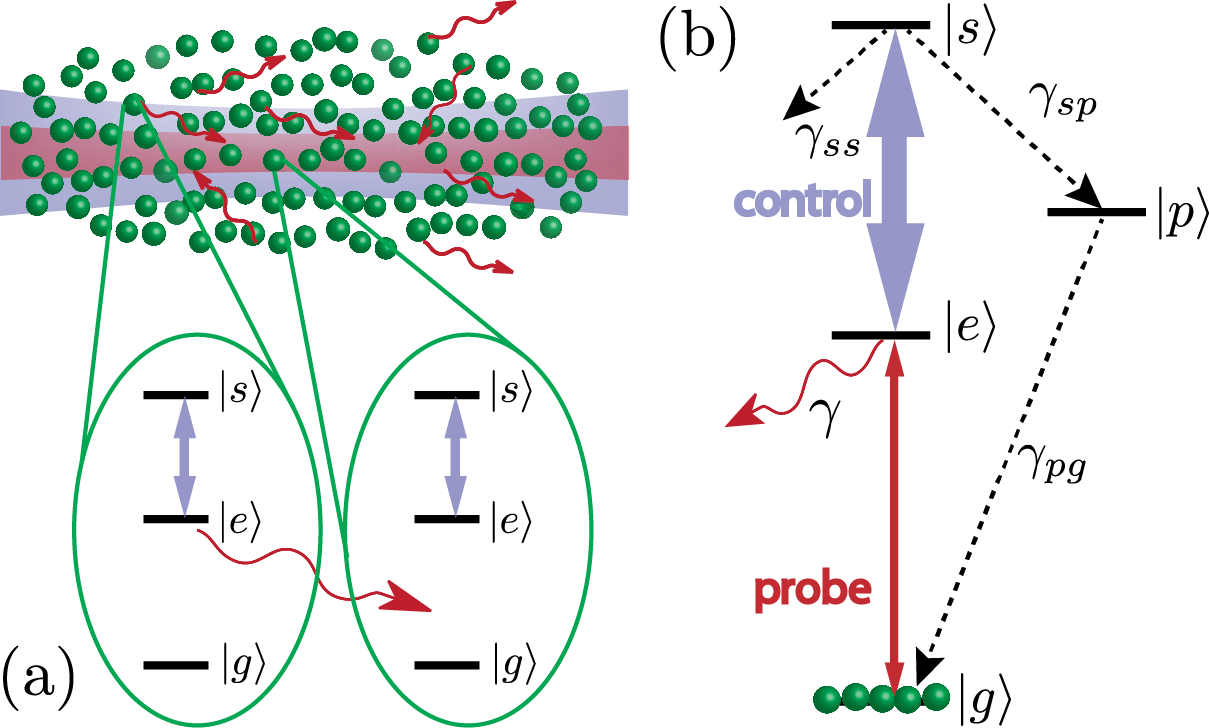}
\caption{\label{fig:pollutant_scheme} 
(a) Illustration of radiation trapping of scattered probe photons in the atomic cloud. Re-absorption of probe photons is possible within the larger control beam. The two three-level atoms that we zoom into schematically represent a process where the left atom emits a photon (red arrow), which is then absorbed by the right atom.
(b) Level scheme for the effective model we introduce to incorporate the pollutant atoms in our numerics. 
}
\end{center}
\end{figure}
While the results of MPS simulations presented in the previous section qualitatively reproduce the experimentally observed effects both in the probe pulse shape and in the steady-state correlation functions, the lack of quantitative agreement suggests that the Rydberg pollutants we register in the experiment may have a strong effect on the probe pulse transmission even at low photon rates $\lesssim 4$ ph$/\mu$s, which is lower than what Figs.~\ref{rates}(b,c) may suggest.
On the other hand, the relatively weak reduction of the test pulse transmission points towards the fact that the Rydberg pollutants have undergone a Rydberg state change and/or that there exists a process that removes them from the path of control and probe beams.
As a possible explanation for the initial source of pollutants, we suggest radiation trapping~\cite{Sadler2017} of scattered probe photons as an initial creation mechanism of Rydberg pollutants, followed by interaction-induced antiblockade and Rydberg-atom~\cite{Tresp2016,Derevianko2015} collisions. 

In this model, the pollutant creation proceeds as follows. Due to the finite extent of the cloud and the large waist of the control beam, photons scattered out of the probe mode do not necessarily escape the medium, but can instead be reabsorbed in state $\Ket{s}$. 
Indeed, we estimate that our atomic cloud has a transverse optical depth of $\sim 13$ at its center, and given that optical depth is a rough estimate of how many times a photon is scattered before leaving the medium, we expect that the lifetime of scattered photons could be enhanced by a factor of order 10. 
This radiation trapping leads to additional atoms in $\Ket{s}$ that are not part of polaritons propagating in the probe direction, but are however able to block the probe photon transmission. 
This effect in itself is not sufficient to explain the observation of ions, as even taking into account radiation trapping, such $\Ket{s}$ state excitations would still be expected to exit the system before the ionization pulse. 
Instead, through this process, atoms in $\Ket{s}$ with all possible angles between pairs of them are created. In this situation, both state-changing Rydberg collisions as well as direct anti-blockade excitation of other Rydberg levels can occur on the microsecond time scale of the experiments~\cite{Tresp2016,Derevianko2015}. 
Atoms in these additional states are not coupled to the control light and therefore are not de-pumped. Summarizing, the radiation trapping gives rise to both (a) the creation of the pollution atoms in the $\ket{{s}}$-state [which ultimately leave the medium] and (b) the creation of \textit{stationary} pollutant Rydberg states (other than state $\ket{s}$), which we observe as ions after field ionization [Fig.~\ref{pulse}(c)].  

\subsection{Effective pollutant model \label{toy_model}}

Simulation of the full pollution process discussed above is prohibitively difficult. Specifically, the MPS model that we have used is only efficient in describing one-dimensional propagation. Treatments of the full scattering problem in three-dimensions, so that radiation trapping is fully accounted for, are possible but currently only at the level of one or two total atomic excitations in the the system~\cite{Schilder2016,Bromley2016,Jennewein2016a}. 
Furthermore, taking into account the full family of Rydberg states and interactions would lead to an explosion of the computational Hilbert space.

Instead, we develop here a toy model that includes the basic features of the pollution process. We do so by modifying the existing MPS model to include an additional atomic 'pollutant' state $\ket{p}$. 
This state is populated by the decay from state $\ket{s}$ at rate $\gamma_{sp}$ as shown in Fig.~\ref{fig:pollutant_scheme} and is assumed to induce the same Rydberg blockade as atoms in state $\ket{s}$ (in future work, it may be interesting to consider extensions where states $\ket{p}$ and $\ket{s}$ have different blockade radii~\cite{Young2018}). 
The population of state $\ket{p}$ can then decay back to the ground state at rate $\gamma_{pg}$. While state $\ket{p}$ is not meant to represent any specific Rydberg state, we take it as a proxy for the pollution process. 
Atoms in state $\ket{p}$ could represent atoms in state $\ket{s}$ that are radiation trapped outside of the probe beam [but still inside the control beam] or atoms that have changed to new Rydberg states that still have a similar blockade radius. 
The decay $\gamma_{pg}$ then takes into account two phenomena: (a) a final escape from the control field of the radiatively-trapped $\ket{s}$ excitations and (b) decay of the other Rydberg pollutant states [note that since the population of the ground state in the MPS model is essentially arbitrary, this decay can also represent decay to other long-lived Rydberg states that interact with $\ket{s}$ only weakly].

In Fig.~\ref{fig:toy_model}(a)-(b), we show  time traces generated by the MPS model with and without pollution for two different input photon rates and compare these time traces to the experimental data. Choosing decay rates of $\gamma_{sp}/2\pi = \gamma_{pg} /2\pi= 100$\,kHz, the modified MPS model provides much closer agreement with the experimental data than  the original simple MPS model.
Despite the simplicity of this toy model, we see that it can explain the much larger initial humps seen in the experimental time traces. Furthermore, in the $g^{(2)}$ correlation function shown in Fig.~\ref{fig:toy_model}(c)-(d), the addition of pollutants to the theory also increases the size of the hump, however in this case for the parameters chosen the hump becomes larger than that seen in the experiment.

\begin{figure}
\centering
\includegraphics{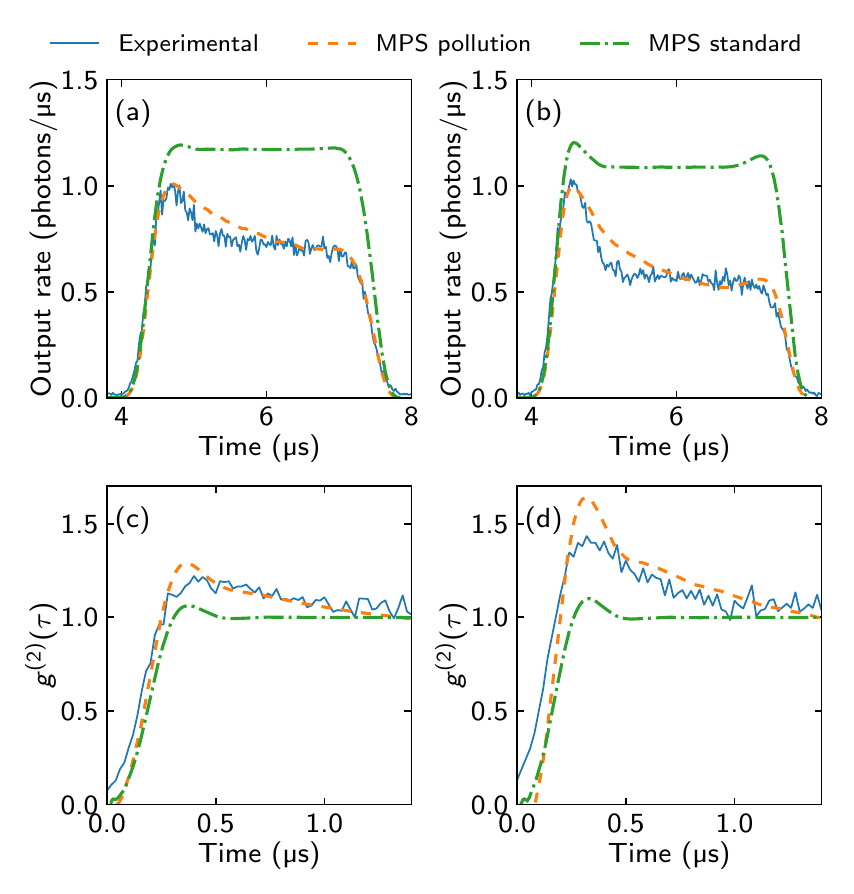}
\caption{\label{fig:toy_model} 
(a)-(b) Intensity output time traces from the experiment,  MPS pollution model, and standard MPS [depicted by blue-solid, orange-dashed and green-dot-dashed lines, respectively], for input rates of (a) 4.2\,ph/$\mu$s = $0.55/\tau_b$ and (b) 7.1\,ph/$\mu$s = $0.92/\tau_b$. The pollution toy model with $\gamma_{sp}/ 2\pi = \gamma_{pg}/ 2\pi = 100$\,kHz shows better agreement with the experimental time traces than the original MPS model with $\gamma_{ss} / 2\pi=40$\,kHz. Correlation function $g^{(2)}$ for (c) 3.0\,ph/$\mu$s = $0.39/\tau_b$ and (d) 7.7\,ph/$\mu$s = $1/\tau_b$ for the toy model with pollution decay rates as above, compared with experiment and the original MPS model with $\gamma_{ss}/ 2\pi = 120$\,kHz. 
All other parameters as in the rest of this manuscript.
The Rydberg interaction is modeled by three exponentials as described in Appendix~\ref{ap:expon_approx}. MPS density matrix simulations used time step $0.01/\gamma$, $N = 60$, and bond dimensions (a) (original) $D = 100$, (pollution) $D = 120$, (b) (original) $D = 140$, (pollution) $D = 180$, (c) (original) $D = 140$, (pollution) $D = 200$, (d) (original) $D = 180$, (pollution) $D = 260$.}
\end{figure}

The success of this toy model leads us to conclude that pollutants do indeed play a major role in the observed output field, and may be the dominant determiner of the size of the humps we see both in the time traces and in $g^{(2)}$. Meanwhile, this simple model neglects effects that are likely present in the system, such as the potential intensity dependence of $\gamma_{sp}$ and $\gamma_{pg}$. The description of such effects requires a deeper understanding of which Rydberg processes take place and lead to pollution. Given the importance of these pollution effects at high intensity, we hope that this work will motivate further experimental and theoretical studies of this phenomenon and how it may be controlled and harnessed for applications.

\section{Conclusions and Outlook \label{sec:outlook}}
In this paper, we have discussed the physics of transmission of photons at high intensities through a Rydberg medium under the conditions of electromagnetically induced transparency. We have utilized a phenomenological model that produces reasonably good qualitative predictions for the time trace of the output  intensities as well as for the steady state output rate. In addition, we utilized numerical MPS techniques to obtain a quantitative simulation of the system. The results of the two theoretical models qualitatively agree with each other. The discrepancy between these simulations and the observed experimental data points to the presence of pollutants. We extend the MPS model to include a simple treatment of pollutants consisting of an additional level. We tune this model to provide a better match to the experimental results. Our work motivates further investigation of high intensity rEIT. It highlights the importance of the role pollutants play in this strongly interacting many-body system, a role that requires additional theoretical and experimental studies and that may eventually be harnessed for applications.


\begin{acknowledgments}
We would like to thank Mari-Carmen Bañuls, Andrew Daley, Eduardo Mascarenhas, Callum Murray, Jeff Thompson, Thomas Pohl 
for stimulating discussions. We acknowledge funding by ARL CDQI, NSF PFC at JQI, the German Research Foundation through Emmy-Noether-grant HO 4787/1-1 and GiRyd project HO 4787/3-1, the Ministry of Science, Research and the Arts of Baden-W\"{u}rttemberg (RiSC grant 33-7533.-30-10/37/1), ERC Consolidator grant RYD-QNLO (grant N. 771417), AFOSR, ARO MURI, ARO, NSF QIS, NSF Ideas Lab, Fundacio Privada Cellex Barcelona, the CERCA Programme/Generalitat de Catalunya, the MINECO Ramon y Cajal Program, the Spanish Ministry of Economy and Competitiveness, through the Severo Ochoa Programme for Centres of Excellence in R\&D (SEV-2015-0522) and Plan Nacional Grant CANS, the Marie Curie Career Integration Grant ATOMNANO, the ERC Starting Grant FoQAL
and the US MURI Grants QOMAND and Grant Photonic Quantum Matter.
EZ acknowledges funding from the Carlsberg Foundation. APM acknowledges support from  UNAM-PAPIIT IA101718 RA101718. PT is supported by the NRC postdoctoral scholarship.
\end{acknowledgments}

\bibliography{library}


\appendix
\section{Extensions of the hard-sphere model} \label{sec:Emil_theory}
%
In this Appendix, we first introduce the technical details of the hard-sphere model discussed in Ref. \parajcite{Zeuthen2017} and then present extensions to this model. In this model, we approximate the interaction between Rydberg polaritons by a hard-sphere potential of radius $r_b$. The projective nature of this interaction means that the resulting many-body Rydberg wavefunction takes a relatively simple form in the position-space representation. We also assume that the polaritons move with a constant velocity $v_g$ in the Rydberg medium. This allows us to use position and time interchangeably. We will use a time basis to denote the position of the $i^{\rm th}$ polariton. For example, a polariton denoted by $t_i$ was created at the beginning of the medium, $r=0$, at time $t_i$. At any time $t>t_i$, the position of the polariton is given by $r_i(t) = (t-t_i)v_g$. 
A many-body pure state of $R$ Rydberg polaritons is described by a time-ordered set of coordinates,
\begin{equation}
|\boldsymbol{t}_R\rangle=|t_1,t_2\cdots t_R\rangle,\ \ t_1<t_2\cdots <t_R.
\end{equation}
We can now define a general density matrix describing this system of polaritons,
\begin{equation}
\rho=\sum_{R,R'}\int\mathcal{D}t\ \mathcal{D}t'\,e_{R,R'}\left(\{\boldsymbol{t}_{R};\boldsymbol{t}_{R'}^{\prime}\}\right)\,
|\boldsymbol{t}_{R'}^{\prime}\rangle\langle\boldsymbol{t}_{R}|,\label{eq:density-matrix-def}
\end{equation}
where $e_{R,R'}\left(\{\boldsymbol{t}_{R};\boldsymbol{t}_{R'}^{\prime}\}\right)$ are the elements of the density matrix, and we have defined the shorthand for the time-ordered integral $\int\mathcal{D}t\ \mathcal{D}t' \equiv \int_{t_{1}<t_{2}\cdots<t_{n}}\prod_{i}dt_{i}\ \int_{t'_{1}<t'_{2}\cdots<t'_{n'}}\prod_{i'}dt'_{i'}$. Note that the density matrix we consider here is the one that results after the entire pulse has entered the medium.

The expression for the density matrix obtained in the hard-sphere model is a generalization of that derived in Ref.~\cite{Gorshkov2013}. In Ref.~\cite{Gorshkov2013}, the authors develop a master-equation-type approach to calculate the outgoing photonic density matrix in the limit that the incident photons, after EIT compression, fit within the medium and that the full medium is blockaded. While the first photon forms a polariton, the subsequent photons get scattered and project the wavefunction of the first polariton. 
Given that the scattered photons are not detected (and hence traced out in our formalism), the density matrix of the outgoing single photon is no longer pure. 
In fact, the coherence in the single-polariton density matrix is related to the timing of the scattering. Ref.~\cite{Zeuthen2017} generalized Ref.~\cite{Gorshkov2013} to the case where the pulse size is larger than the blockade radius. 
We now provide an intuition for the output density matrix under the assumptions of the hard-sphere model. 
The density matrix consists of  coherences between many-polariton states. 
The polaritons must be at least one blockade time $\tau_b$ apart from each other. Let $I(\boldsymbol{t}_R)$ denote the region $\bigcup_{i=1}^R [t_i,t_i+\tau_b)$ in which incoming photons are scattered by a polariton state. 
The quantum coherence (developed between the polariton states $|\boldsymbol{t}_R\rangle$ and $|\boldsymbol{t}^{\prime}_{R'}\rangle$) \changed{arises from the fact that the projections of the Rydberg polariton wave function associated with a given set of scattering events does not fully determine the position of the polaritons. Put slightly differently, we can regard the wave function of the incoming light as a coherent superposition of arrival times for the photons; those coherences that are not destroyed by a given set of scattering events are then simply mapped into coherences of the Rydberg wave function. Since we assume that the scattered photons are not detected, they are traced out in our theory. Hence, we must perform an (incoherent) integral over all sets of scattering times.}{comes from the scattering events whose time is the same 
for both states.} 
To this end, let us represent the temporal region \changed{in which scattering events can take place without destroying the coherence between states $|\boldsymbol{t}_R\rangle$ and $|\boldsymbol{t}^{\prime}_{R'}\rangle$ as}{where simultaneous scattering is possible as} $I(\boldsymbol{t}_R)\cap I(\boldsymbol{t}^{\prime}_{R'})$. For $m$ scatterings, the coherence factor becomes  $\left(\int_{I(\boldsymbol{t}_R)\cap I(\boldsymbol{t}^{\prime}_{R'})}h^{2}(\tau)d\tau\right)^m$ in a straightforward generalization of Ref.~\cite{Gorshkov2013}.  Utilizing the above insight, the elements of the density matrix for a Fock state input, $|\psi_{\rm in}\rangle=|n_{\rm in}\rangle$, can be expressed as 
\begin{widetext}
\begin{equation}
e_{R,R'}\left(\{\boldsymbol{t}_{R};\boldsymbol{t}^{\prime}_{R'}\}\right)	=\delta_{R,R'}\prod_{i,i'}\Theta\left(t_{i+1}-t_{i}-\tau_{b}\right)\Theta\left(t_{i'+1}^{\prime}-t_{i'}^{\prime}-\tau_{b}\right)h(t_{i})h(t_{i'}^{\prime}) \frac{n_{\rm in}!}{(n_{\rm in}-R)!}\left(\int_{I(\boldsymbol{t}_R)\cap I(\boldsymbol{t}^{\prime}_{R'})}h^{2}(\tau)d\tau\right)^{n_{\rm in}-R}, 
\label{diagonal_eRR}
\end{equation}
where the factor $\frac{n_{\rm in}!}{(n_{\rm in}-R)!}$  is the number of ways in which one can pick an ordered set of R elements (polaritons) out of the $n_{\rm in}$ incoming photons. Note that, for an input state with a definite photon number, it is not possible to have any coherence in the output between states with a different number of polaritons $R\neq R'$ since the environment knows the number of scattered photons and thus the number of remaining polaritons. Now we can generalize the result for an input coherent state $|\alpha\rangle=\sum_n\frac{\alpha^{n}}{\sqrt{n!}}e^{-|\alpha|^{2}/2}|n\rangle$, where for simplicity we assume $\alpha \in \mathbb{R}_+$. The general density matrix element is given by
\begin{equation}
e_{R,R'}\left(\{\boldsymbol{t}_{R};\boldsymbol{t}^{\prime}_{R'}\}\right)	=\prod_{i,i'}\Theta\left(t_{i+1}-t_{i}-\tau_{b}\right)\Theta\left(t_{i'+1}^{\prime}-t_{i'}^{\prime}-\tau_{b}\right)h(t_{i})h(t_{i'}^{\prime})e^{-\alpha^{2}}\alpha^{\left(R+R'\right)}e^{\alpha^{2}\int_{I(\boldsymbol{t}_R)\cap I(\boldsymbol{t}^{\prime}_{R'})}h^{2}(\tau)d\tau}.
\label{general_eRR}
\end{equation}
We can now utilize the expression for the general density matrix element to derive expressions for correlation functions such as $G^{(1)}(t,t')$.

\subsection{Expressions for $G^{(1)}(t,t)$ and $G^{(1)}(t,t')$ for time-varying $h(t)$}
Here we present the expressions, used to plot Fig.~\ref{MPSuniform} in the main text, for  $G^{(1)}(t,t)$ and $G^{(1)}(t,t')$ for time-varying $h(t)$.
By definition, we know that 
\beqa
G^{(1)}(t,t)&=&\text{Tr}\left[\mathcal{E^{\dagger}}(t)\mathcal{E}(t)\rho\right] = 
\sum_{R} \int\mathcal{D} t\ \  e_{R,R}\left(\{\boldsymbol{t}_{R};\boldsymbol{t}_{R}\}\right)\sum_{i=1}^R \delta(t-t_{i}).
\label{eq:G1-def-intermediate}
\eeqa
The above expression can be simplified further by carrying out the integrals over all $t_i>t$.
Using Eqs.\ \eqref{general_eRR} and \eqref{eq:G1-def-intermediate} and assuming that  the rise time of the pulse $t_{\rm rise}$ fits at most $R_{r} $ [note that always $R_{r}\geq 1$] polaritons, and that the pulse begins at $t=0$, we arrive at  
\beqa
G_{R_{r}}^{(1)}(t,t)&=&
|f(t)|^{2}\sum_{R=1}^{\lceil t/\tau_{b}\rceil}\exp[R_{\rs in}(R-R_c-1)\tau_{b}]R_{\rs in}^{R}
\left(\prod_{j=1}^{j=R_c}
\int_{t_{j-1}+\tau_{b}}^{t-(R-j)\tau_{b}}dt_{j}|f(t_{j})|^{2}	\right)\nn\\
&&
\hspace{0.7in}\times\exp\left[R_{\rs in}\sum_{i=1}^{R_c}(F(t_{i}+\tau_{b})-F(t_{i}))\right]
\frac{\left(t-t_{R_c}-(R-R_c)\tau_{b}\right)^{R-R_c-1}}{(R-R_c-1)!}\exp[-R_{\rs in}F(t)],
\label{GRcDiagonal}
\eeqa
where we used $R_{\rs in}$, $R_c$, $t_0$, $f$, and $F$ defined as $R_{\rs in}=\alpha^2 \bar{h}^2,$ $R_c=\min[R_r,R-1]$, $t_0=-\tau_b$, $f(t)=h(t)/\bar{h}$, and $F(t)=\int_{0}^{t}|f(t)|^{2}$, respectively. 
The amplitude $\bar{h}$ is defined as the (constant) amplitude of the incoming photon for times greater than $t_{\rs rise}$. Notice that we do not include the fall time of the Tukey pulse in this model. 

Analogously, using Eq.~\eqref{general_eRR}, we can calculate 
the off-diagonal $G^{(1)}(t,t')$. To this end, we introduce $t_>=\max(t,t')$ and  $t_<=\min(t,t')$. Then the expression for $G^{(1)}(t,t')$ for $t_{>}-t_{<}<\tau_{b}$
takes the form 
\beqa
G_{R_{r}}^{(1)}(t,t')&=&
\frac{f(t_>)}{f(t_<)}
G_{R_{r}}^{(1)}(t_{<},t_{<})\exp\left[-R_{\rs in}(-F(t_{<})+F(t_{>}+\tau_{b})-F(t_{<}+\tau_{b})+F(t_{>}))\right];
\eeqa
whereas, for $t_{>}-t_{<}>\tau_{b}$,  we obtain 
\beqa
G_{R_{r}}^{(1)}(t,t')&=&
\frac{e^{-R_{\rs in}(F(t_{<}+\tau_{b})+F(t_{>}+\tau_{b})-F(t_{>})-F(t_{<}))}}{f(t)f(t')R_{\text{in}}}G_{R_{r}}^{(1)}\left(t_{<},t_{<}\right)G_{R_r,t_{<}+\tau_{b}}^{(1)}\left(t_{>}-t_{<}-\tau_{b},t_{>}-t_{<}-\tau_{b}\right),
\eeqa
where 
$G_{R_r,t_s}^{(1)}(t,t)$ is defined using Eq.~\eqref{GRcDiagonal} but with $h(t)$ replaced by $h_{t_s}(t)=\Theta(t) h(t+t_s)$ and $f(t),F(t)$ are defined using $h_{t_s}(t)$.

\end{widetext}


\section{MPS treatment of Rydberg EIT}\label{sec:MPSappendix}
Light propagation though atomic ensembles in the high-intensity limit is a difficult problem to study numerically, as we need to describe a driven-dissipative system in the regime where many-body correlations are important. To do so, we extend a recently developed technique~\cite{Manzoni2017} that is based on mapping light propagation to the physics of a driven-dissipative spin chain and then solving the spin-chain dynamics using the matrix product ansatz~\cite{Verstraete2008,Schollwock2011}. Here we briefly review this technique while referring the reader to Ref.~\cite{Manzoni2017} for further details.

In quasi-1D light propagation experiments, such as the one presented here, the standard approach is to study the paraxial Maxwell-Bloch equations. Instead, in our MPS approach we take advantage of a mapping of these equations to the dynamics of a chain of atoms. The atoms couple via a dipole transition $\ket{g}$-$\ket{e}$ to the quantum light field $E(z,t)$ with central wavevector $k$ propagating in the $z$-direction, and at any point the resulting field is just the sum of the input field and the field generated by the $M$ atomic dipoles. This yields the generalized input-output relation~\cite{Caneva2015,Manzoni2017} for the electric field 
\begin{equation}\label{eq:spinlight}
E(z,t)  = E_\text{in}(z,t) + i\sum_{j=1}^M \sqrt{\frac{\Gamma_j}{2}}e^{i k |z - z_j|}\sigma^j_\text{ge}(t),
\end{equation}
where $\Gamma_j$ describes the strength of the coupling between the paraxial input mode and the atom $j$ described by $\sigma^j_\text{ge}$ defined as $\ket{g}_j\hspace{-.3em}\bra{e}_j$.

This field then couples back to the atoms, where the coupling of the atoms to the input field is given by $H_\text{drive} = -\sum_{j=1}^M\sqrt{\Gamma_j/2}\left(E_\mathrm{in}(t,z)\sigma^j_\text{eg}+\text{H.c.}\right)$. Coherent input $E_\mathrm{in}(t,z) = \mathcal{E}_\mathrm{in}(t)e^{i k z}$, as used in our experiment, can be treated as a classical field without approximation~\cite{Mollow1975}. Furthermore, the field generated by one atom may then couple to another atom giving an effective dipole-dipole interaction between the atoms~\cite{Caneva2015,Manzoni2017}, 
\begin{equation}\label{eq:Hdd_1d}
H_\mathrm{dd}=\sum_{j,l=1}^M \frac{\sqrt{\Gamma_j\Gamma_l}}{2}\sin(k|z_j-z_l|)\sigma_\text{eg}^{j}\sigma_\text{ge}^{l}. 
\end{equation}
Dissipation is also present as photons leave the ensemble in the chosen paraxial mode. This dissipation is described by a Lindbladian
\begin{multline}\label{eq:diss_1d}
\mathcal{L}_\mathrm{dd}[\rho]= \sum_{j,l=1}^M\frac{\sqrt{\Gamma_j\Gamma_l}}{2}\cos(k|z_j-z_l|)\\\times(2\sigma_\text{ge}^{j}\rho\sigma_\text{eg}^{l}-\sigma_\text{eg}^{j}\sigma_\text{ge}^{l}\rho-\rho\sigma_\text{eg}^{j}\sigma_\text{ge}^{l}). 
\end{multline}
Photons can also leave via spontaneous emission at rate $\gamma$ into other modes (taken to be the rate of  spontaneous emission of a single atom in free space). This process corresponds to the Lindbladian
\begin{equation}
\mathcal{L}_\text{spont}[\rho]= \frac{\gamma}{2}\sum_{j=1}^M(2\sigma_\text{ge}^{j}\rho\sigma_\text{eg}^{j}-\sigma_\text{eg}^{j}\sigma_\text{ge}^{j}\rho-\rho\sigma_\text{eg}^{j}\sigma_\text{ge}^{j}).
\end{equation}


The dynamics of this driven spin system can then be solved to yield the output field after propagation through the ensemble using Eq.~\eqref{eq:spinlight}. While above we have only discussed the coupling of the light to the transition $\ket{g}$-$\ket{e}$, additional atomic dynamics, such as those due to the presence of Rydberg levels and due to the dipolar interaction between them, may be added to the master equation without altering the above results.

Solving the spin dynamics itself is challenging. Indeed, in the experiment, the atomic cloud contains tens of thousand of atoms, a number that cannot be feasibly modeled numerically. The dynamics may be found by evolving directly the density matrix in time or using the quantum jump formalism~\cite{Molmer1993} for up to $\sim20$ atoms. To go beyond this, we first recognize that, in many propagation experiments, the number of atoms present is not of primary importance. Instead, the important quantity is the number of atoms multiplied by their coupling strength to the probe beam. The original large number of atoms in the ensemble can then be modeled by a much smaller chain of atoms where the overall optical depth of the system and optical depth per Rydberg blockade radius are conserved. Specifically, we can divide the atom cloud along the $z$ axis into $N$ slices of width $a$ centered at positions $z_l$. Then the
collection of atoms within one slice $\sum_{|z_j - z_l|< a/2}\sqrt{\Gamma_j/2}\sigma^j_{ge}$ may be replaced by an effective atom $\sqrt{\Gamma_l/2}\sigma^l_{ge}$, whose coupling to the propagating field is the sum of the couplings of the original atoms $\Gamma_l = \sum_{|z_j - z_l|< a/2}\Gamma_j$. In this way, the optical properties of the system are preserved as long as single-effective-atom saturation effects are not present. To avoid saturation, the number of slices must be kept sufficiently large, which can be checked in numerical simulations by verifying that observables are invariant under changes in $N$ provided that $\OD$ and $\OD_b$ are maintained constant (see Appendix~\ref{app:conv_bd_and_ns}). Grouping the atoms in this way also allows the non-uniform distribution of atoms in the cloud to be conveniently modeled by a non-uniform coupling to each atomic slice.

By reducing the full propagation dynamics to a model of tens of atoms, the dynamics can now be solved using matrix product states. There are then two possible treatments: (1) to represent the state of the system as a pure state MPS and propagate using the quantum jump formalism~\cite{Daley2009a,Daley2014a} or quantum state diffusion~\cite{Gisin1992}, or (2) to convert the density matrix of the system to an MPS and use the Liouvillian superoperator approach to find the time dynamics~\cite{Zwolak2004,Verstraete2004}. In Ref.~\cite{Manzoni2017}, light propagation was studied exclusively using the quantum jump method, and here we use that method to find the steady-state output shown in Fig.~\ref{fig:comp_exp_MPS}(d).

On the other hand, we find that using quantum jump trajectories to find the time traces in Fig.~\ref{fig:comp_exp_MPS}(a)-(c) is inefficient, as is quantum state diffusion, due to the large number of trajectories needed to reduce statistical noise. Instead, we propagate the density matrix using the master equation $\partial_t \rho = \mathcal{L}(\rho)$ ($= -i[H_\text{dd} + H_\text{drive},\rho] + \mathcal{L}_\text{dd}(\rho) + \mathcal{L}_\text{spont}(\rho)$ for the simple spin model above). To do so, we first map the density matrix $\rho$ to a vector~\cite{Zwolak2004} by identifying local vector basis states, e.g., for two-level atoms $\{\ket{g}\bra{g},\ket{g}\bra{e},\ket{e}\bra{g},\ket{e}\bra{e}\}\rightarrow \{|gg), |ge),|eg),|ee)\}$. In this basis, the density operator can be rewritten as an MPS $|\rho) = \sum_{\beta_1,\ldots,\beta_N}A^{\beta_1}\cdots A^{\beta_N}|\beta_1)\ldots|\beta_N)$, where the sum is over the basis states $|\beta_j)$ for each atom $j$. MPS states are generalizations of product states, e.g., $|\rho) = \sum_{\beta_1,\ldots,\beta_N}c^{\beta_1}\cdots c^{\beta_N}|\beta_1)\ldots|\beta_N)$, where instead of having a complex coefficient $c^{\beta_j}$ associated with the state of each atom we have a matrix $A^{\beta_j}$. This allows for entanglement to be introduced into the state in a controlled way by increasing the size of the matrices associated with each site. 

Operators acting to the left or right of $\rho$, as required to represent the Liouvillian, can also be mapped into our new vector space using Kronecker products. Operator products such as $O^j\rho I^j$, where $O^j$ and $I^j$ are the matrix representations in the original basis of operator $O$ and the identity acting at site $j$, become $O^j\tilde\otimes I^j|\rho)$. Here we have defined the Kronecker product $A\tilde\otimes B = A\otimes B^T$ for notational convenience. 

\begin{widetext}
In this way, the Liouvillian for the spin-model described above becomes
\begin{multline}
\mathcal{L} = \sum_{l>j}\sqrt{\frac{\Gamma_j\Gamma_l}{4}}\left\{e^{ika(l-j)}\left[(I^j\tilde\otimes\sigma_{eg}^j-\sigma_{eg}^j\tilde\otimes I^j)\sigma_{ge}^l\tilde\otimes I^l + \sigma_{ge}^j\tilde\otimes I^j(I^l\tilde\otimes \sigma_{eg}^l-\sigma_{eg}^l\tilde\otimes I^l)\right]+ \right.\\
\left.e^{-ika(l-j)}\left[I^j\tilde\otimes\sigma_{eg}^j(\sigma_{ge}^l\tilde\otimes I^l-I^l\tilde\otimes \sigma_{ge}^l) + (\sigma_{ge}^j\tilde\otimes I^j-I^j\tilde\otimes\sigma_{ge}^j)I^l\tilde\otimes \sigma_{eg}^l\right]\right\} + \sum_j L_j,
\end{multline}
where
\begin{equation}\label{eq:L_local}
 L_j =  \frac{\Gamma_j + \gamma}{2}\left(2\sigmạ^j_{ge}\tilde\otimes\sigmạ^j_{eg}-\sigma_{ee}^j\tilde\otimes I^j-  I^j\tilde\otimes\sigma_{ee}^j\right)
+ i\sqrt{\frac{\Gamma_j}{2}}\mathcal{E}_\mathrm{in}(t)\left[e^{i k z_j}(\sigma_{eg}^j\tilde\otimes I^j - I^j\tilde\otimes\sigma_{eg}^j)+e^{-i k z_j}(\sigma_{ge}^j\tilde\otimes I^j - I^j\tilde\otimes\sigma_{ge}^j)\right].
\end{equation}
We can then express the entire Liouvillian as a matrix product operator with site matrices given by
\begin{equation}
\mathcal{L}_j =\left(
\scriptsize\begin{array}{cccccc}
  I^j\tilde\otimes I^j & \sqrt{\frac{\Gamma_j}{2}}e^{ika}(I^j\tilde\otimes\sigma_{eg}^j-\sigma_{eg}^j\tilde\otimes I^j)  & \sqrt{\frac{\Gamma_j}{2}}e^{ika}\sigma_{ge}^j\tilde\otimes I^j &  \sqrt{\frac{\Gamma_j}{2}}e^{-ika}I^j\tilde\otimes\sigma_{eg}^j & \sqrt{\frac{\Gamma_j}{2}}e^{-ika}(\sigma_{ge}^j\tilde\otimes I^j-I^j\tilde\otimes\sigma_{ge}^j) & L_j \\
  0		 & e^{ika}I^j\tilde\otimes I^j    & 0				& 0			  & 0			    & \sqrt{\frac{\Gamma_j}{2}}\sigma_{ge}^j\tilde\otimes I^j\\
  0		 & 0			    & e^{ika}I^j\tilde\otimes I^j	& 0			  & 0			    & \sqrt{\frac{\Gamma_j}{2}}(I^j\tilde\otimes \sigma_{eg}^j-\sigma_{eg}^j\tilde\otimes I^j)\\
  0		 & 0			    & 0				& e^{-ika}I^j\tilde\otimes I^j  & 0			    & \sqrt{\frac{\Gamma_j}{2}}(\sigma_{ge}^j\tilde\otimes I^j-I^j\tilde\otimes \sigma_{ge}^j)\\
  0		 & 0			    & 0				& 0			  & e^{-ika}I^j\tilde\otimes I^j  & \sqrt{\frac{\Gamma_j}{2}}I^j\tilde\otimes \sigma_{eg}^j\\
  0		 & 0			    & 0				& 0			  & 0			    & I^j\tilde\otimes I^j\\
  \end{array}
\right)
\end{equation}
for $j =  2, \ldots, N-1$, and
\begin{align}
\mathcal{L}_1 & = \left(I^1\tilde\otimes I^1, \sqrt{\frac{\Gamma_1}{2}}e^{ika}(I^1\tilde\otimes\sigma_{eg}^1-\sigma_{eg}^1\tilde\otimes I^1), \sqrt{\frac{\Gamma_1}{2}}e^{ika}\sigma_{ge}^1\tilde\otimes I^1,  \sqrt{\frac{\Gamma_1}{2}}e^{-ika}I^1\tilde\otimes\sigma_{eg}^1, \sqrt{\frac{\Gamma_1}{2}}e^{-ika}(\sigma_{ge}^1\tilde\otimes I^1-I^1\tilde\otimes\sigma_{ge}^1), L_1\right),\\
\mathcal{L}_N & = \left(L_N,\sqrt{\frac{\Gamma_N}{2}}\sigma_{ge}^N\tilde\otimes I^N,\sqrt{\frac{\Gamma_N}{2}}(I^N\tilde\otimes \sigma_{eg}^N-\sigma_{eg}^N\tilde\otimes I^N),\sqrt{\frac{\Gamma_N}{2}}(\sigma_{ge}^N\tilde\otimes I^N-I^N\tilde\otimes \sigma_{ge}^N),\sqrt{\frac{\Gamma_N}{2}}I^N\tilde\otimes \sigma_{eg}^N,I^N\tilde\otimes I^N\right)^T.
\end{align}

\end{widetext}

The above operators can now be applied to the MPS representing the density operator to evolve the system in time. For example, a linear expansion of the master equation could be achieved by applying $1+dt\mathcal{L}$ to $|\rho)$ for sufficiently small time step $dt$, where $1+dt\mathcal{L}$ is found from the matrix product operator above by simply multiplying all rates $\Gamma_j,\gamma, \sqrt{\Gamma_j}\mathcal{E}_\text{in}$ by $dt$ and adding $I^j\tilde\otimes I^j/N$ to each $L_j$. Here, instead, to allow for larger time steps, we use a Runge-Kutta 4th order method~\cite{Press:NR}. The steady state may also be found by minimizing $(\rho|\mathcal{L}^\dagger\mathcal{L}|\rho)$ using traditional DMRG algorithms~\cite{Cui2015,Mascarenhas2015}, however we find that time evolution in the quantum jump formalism yields the steady state more efficiently for the problem at hand. In this new vector space, we use the identity $tr(A^\dagger \rho) = ( A|\rho)$ to calculate expectation values, where $|A)$ is the MPS vector mapping of the operator $A$. 

\subsection{Approximation of Rydberg power-law interactions by a series of exponentials}\label{ap:expon_approx}

The MPO presented above describes the interaction and propagation of light in a 1D channel. We now show how this model is extended to include the case where the atoms also have a third level $|s\rangle$ with Rydberg interactions of the form $V(r) = C_6/r^6\sum_{j < l}\sigma^j_{ss}\sigma^l_{ss}$.

The classical driving of the control laser, $H_\text{control} = \Omega\sum_j(\sigma_{es} + \sigma_{se})/2$, and decoherence of $\ket{s}$,  $\mathcal{L}_\text{ss}[\rho]= \gamma_{ss}\sum_{j=1}^M(\sigma_\text{ss}^{j}\rho\sigma_\text{ss}^{j}-\sigma_\text{ss}^{j}\sigma_\text{ss}^{j}\rho/2-\rho\sigma_\text{ss}^{j}\sigma_\text{ss}^{j}/2)$, are trivially included in the local part of the Liouvillian [Eq.~\eqref{eq:L_local}]. On the other hand, the interaction term is more complicated, as power-law decays have no known compact MPO representation. 
However, such decays can be approximated to arbitrary precision over finite distances by sums of exponentially decaying interactions of the form $V(r) \approx \sum_j \eta_j \lambda_j^r$~\cite{Pirvu2010, Crosswhite2008, Frowis2010a}. Here we use the technique described in Ref.~\cite{Pirvu2010} to find approximations of the Rydberg interaction over the range appropriate for the atomic cloud used in the experiment using 3-6 exponentials. 
In doing so, we also recognize that the large strength of the Rydberg interaction at short range can lead to numerical instabilities and instead fit the sum of exponentials to a fixed core strength at short range, as shown in Fig.~\ref{fig:conv_exp}(a). This is justified, as above a certain value the Rydberg interaction detunes the 
$|s\rangle$ levels to the extent they no longer play a role in the physics.

\begin{figure}[b]
\begin{center}
\includegraphics{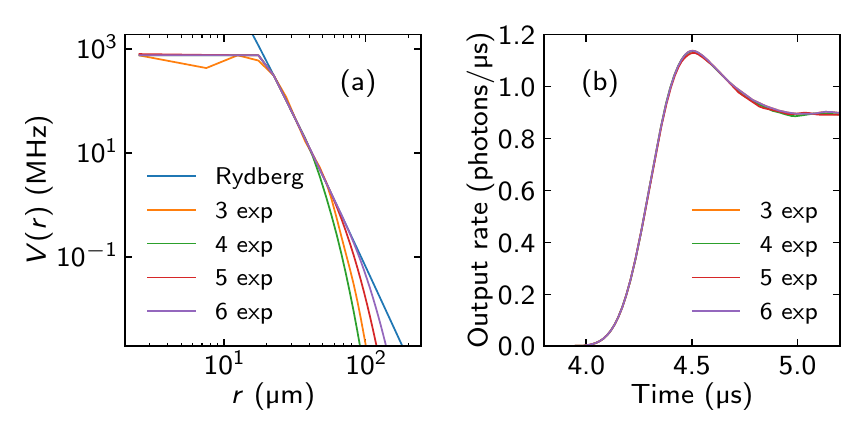}
\caption{(a) The Rydberg interaction is truncated at 
$V = 20 \gamma$ and is approximated using a sum of interactions with exponential form. The specific interaction being modeled has $C_6/ 2\pi =1.8782\times 10^{14}$Hz$\,\mu$m$^{-6}$. Note that we only require a good approximation at the atomic positions, and in between the potential may take arbitrary values. Here we plot the approximate potential only at the positions of the atoms assuming a spin chain with interatomic distance $a = 2.5\,\mu$m. (b) The number of exponentials used leads to only minor differences in the time trace shown here for an input photon rate of 10.4\,ph/$\mu$s. Parameters used in MPS simulations: $\gamma/ 2\pi = 6.065$\,MHz, $\OD = 33$, $\gamma_{ss} / 2\pi= 40$\,kHz, $N = 60$, $D = 100$. The atomic cloud has a Gaussian distribution $n(z) = \exp[-z^2/(2\sigma^2)]$ with $\sigma = 23.6\,\mu$m.}\label{fig:conv_exp}
\end{center}
\end{figure}

In Fig.~\ref{fig:conv_exp}(b), 
we calculate the output intensity given a Tukey function input with incoming photon rate $R_{in}=10.4$\,ph/$\mu$s, and for different approximations of the Rydberg interaction ranging from 3 to 6 exponentials.
No great difference is seen between the curves produced with 3-6 exponentials.
Furthermore, we have also checked that changing the core cutoff from 20$\gamma$ 
(which is shown in the figure) to 30$\gamma$ (not plotted) 
makes no difference to the dynamics.

\begin{figure}[t!]
\begin{center}
\includegraphics[width= 1\columnwidth]{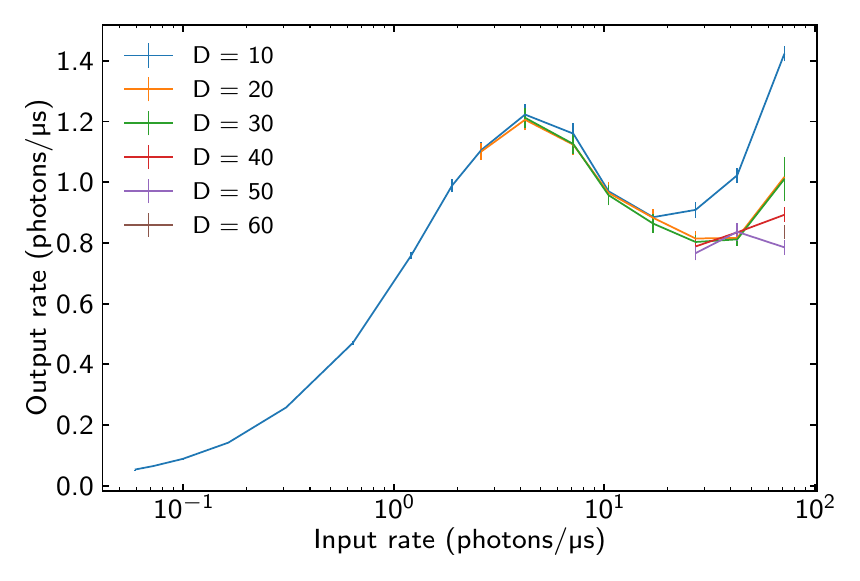}
\caption{Convergence of the predicted steady-state output photon rate with bond dimension $D$ of the MPS used in the quantum jump trajectories. Parameters used in MPS simulations: $\gamma/ 2\pi =  6.065$\,MHz, $\OD = 33$, $\gamma_{ss} / 2\pi=40$\,kHz, $N = 70$. The Rydberg interaction is modeled by five exponentials as described in Appendix~\ref{ap:expon_approx}. The atomic cloud has a Gaussian distribution $n(z) = \exp[-z^2/(2\sigma^2)]$ with $\sigma = 23.6\mu$m.
The vertical bars denote the statistical errors $s$ in the averaging of intensities over all trajectories.}\label{fig:conv_ss}
\end{center}
\end{figure}

\subsection{Convergence with bond dimension and number of spins}\label{app:conv_bd_and_ns}
The accuracy of the MPS methods depends on how well the quantum state of the system can be approximated by an MPS of constrained bond dimension $D$. Furthermore, we have modeled the system consisting of thousands of atoms by tens of atoms. To test that both of these approximations allow the nature of the light propagation to be faithfully represented, we test for convergence of the observable dynamics in both the bond dimension and in the number of atoms used in the simulations.

\begin{figure}[t]
\begin{center}
\includegraphics[width= 0.98\columnwidth]{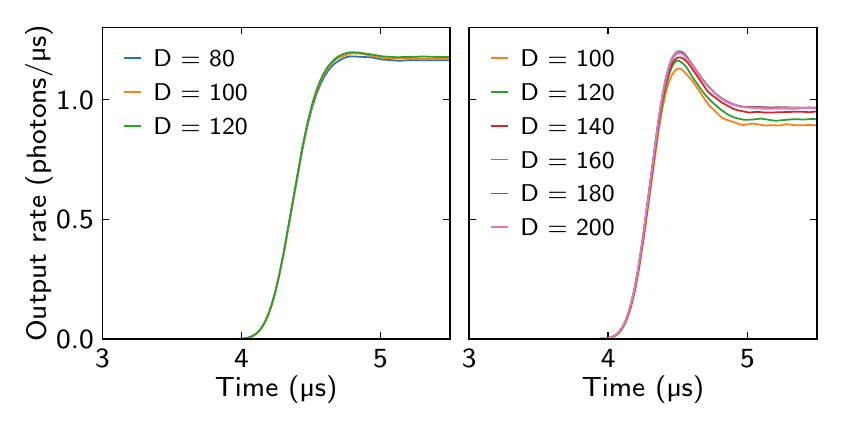}
\caption{Convergence of the time traces with bond dimension of the MPS for the input rate of 4.2\,ph/$\mu$s (left) and 10.4\,ph/$\mu$s (right). Higher input photon rates require larger bond dimension, where for an input rate of 4.2\,ph/$\mu$s the time trace has already converged for $D= 100$, while at input rate of 10.4\,ph/$\mu$s convergence is not seen until $D = 180$. Parameters used in MPS simulations: $\gamma/ 2\pi =  6.065$\,MHz, $\OD = 33$, $\gamma_{ss}/ 2\pi =40$\,kHz, $N = 60$. The Rydberg interaction is modeled by five exponentials as described in Appendix~\ref{ap:expon_approx}. The atomic cloud has a Gaussian distribution $n(z) = \exp[-z^2/(2\sigma^2)]$ with $\sigma = 23.6\mu$m.}\label{fig:conv_time}
\end{center}
\end{figure}

In Fig.~\ref{fig:conv_ss}, we show how increasing the bond dimension of the MPS used in the quantum jump simulations for finding the steady-state affects the observed output steady-state intensity. For input intensities below 10 \,ph/$\mu$s, the output is already well approximated by MPS with $D=10$. For higher intensities, larger bond dimension is required: for the maximum input rate in the experiment of 71.8 ph/$\mu$s, convergence requires $D>50$, suggesting a build up of entanglement at that rate. To calculate the steady-state output, we run quantum jump trajectories under constant input photon flux for a time of $\sim 10000/\gamma$, and after neglecting an initial equilibration time, calculate the average intensity $I_{ss} = \sum^T_j I_j/T$, where $I_j$ are the intensities at the $T$ discrete time steps of the simulation. This averaging is accompanied by statistical error $s = \sqrt{\text{Var}(\sum^T_j I_j/T)} = \sqrt{\sum^T_{j,l} \text{Cov}(I_j, I_l)/T}$ taking into account the correlations in the time series. As shown in the error bars in Fig.~\ref{fig:conv_ss}, this error increases for higher input intensities due to the larger numbers of quantum jumps in the evolution.

Simulations using an MPS representation of the full density matrix require a larger bond dimension for convergence. However, despite this reduction in computational efficiency, this method may still be more efficient that the quantum jump approach as no summation over trajectories is required (where tens of thousands of trajectories are typically required for the convergence of a time trace). In Fig.~\ref{fig:conv_time}, we show the convergence of the time traces for two different input rates. For an input rate of 4.2\,ph/$\mu$s, a bond dimension of 100 is sufficient, while for 10.4\,ph/$\mu$s a bond dimension of 180 is required.
\begin{figure}[h!]
\begin{center}
\includegraphics[width= 1\columnwidth]{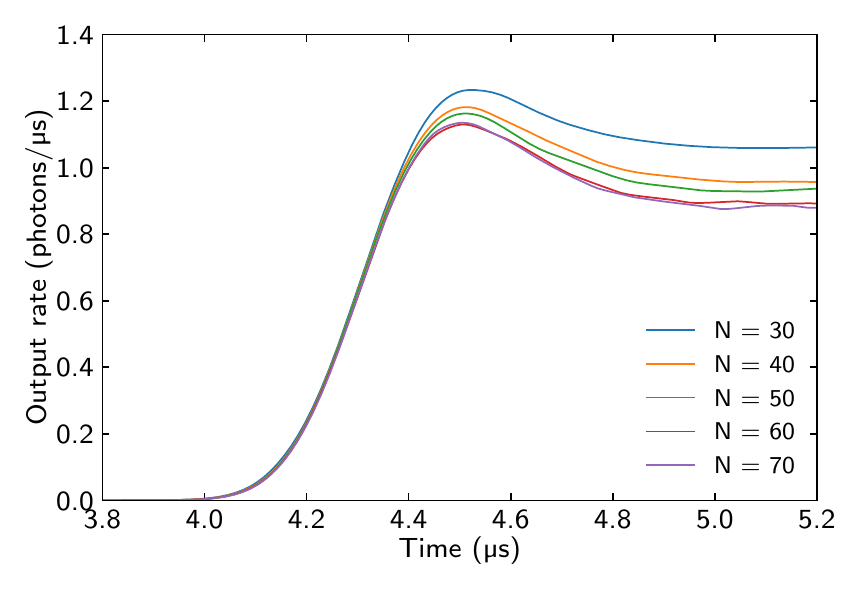}
\caption{Convergence of the time trace of the MPS simulation with the number of effective atoms $N$ used in the spin model. Here we show the time trace for an input rate of 10.4\,ph/$\mu$s. The simulations are all done for maximum bond dimension 100. Parameters used in MPS simulations: $\gamma/2\pi =  6.065$\,MHz, $\OD = 33$, $\gamma_{ss}/2\pi =40$\,kHz. The Rydberg interaction is modeled by five exponentials as described in Appendix~\ref{ap:expon_approx}. The atomic cloud has a Gaussian distribution $n(z) = \exp[-z^2/(2\sigma^2)]$ with $\sigma = 23.6\,\mu$m.}\label{fig:conv_number}
\end{center}
\end{figure}
Finally, in Fig.~\ref{fig:conv_number}, we show the convergence of the time trace for an input rate of 10.4\,ph/$\mu$s with the number of atoms (slices) used in the simulations. For smaller number of atoms, e.g., $N=30$, the time trace shows an overestimate of the output photon rate,
however the qualitative behavior is still present. For higher number of atoms $N = 60$ the time traces converge.

\end{document}